\documentstyle[12pt,epsf,axodraw]{article}
\newcommand {\beq}{\begin{equation}}
\newcommand {\eeq}{\end{equation}}
\newcommand {\beqa}{\begin{eqnarray}}
\newcommand {\eeqa}{\end{eqnarray}}
\newcommand {\beqan}{\begin{eqnarray*}}
\newcommand {\eeqan}{\end{eqnarray*}}
\newcommand {\n}{\nonumber \\}
\newcommand {\eq}[1]{eq.~(\ref{#1})}

\def\mib#1{\mbox{\boldmath $#1$}}
\def\ep{\epsilon}

\begin{document}
\setlength{\oddsidemargin}{0cm}
\setlength{\baselineskip}{7mm}

\begin{titlepage}
 \renewcommand{\thefootnote}{\fnsymbol{footnote}}
$\mbox{ }$
\begin{flushright}
\begin{tabular}{l}
hep-th/9802085\\
KEK-TH-559 \\
TIT/HEP-382\\
\end{tabular}
\end{flushright}

{}~~\\
{}~~\\
{}~~\\

\vspace*{0cm}
    \begin{Large}
       \vspace{2cm}
       \begin{center}
         {Space-Time Structures from IIB Matrix Model}      \\
       \end{center}
    \end{Large}

  \vspace{1cm}

\begin{center}

           Hajime A{\sc oki}$^{1)}$\footnote
           {
e-mail address : haoki@theory.kek.jp, JSPS research fellow},
           Satoshi I{\sc so}$^{1)}$\footnote
           {
e-mail address : iso@theory.kek.jp},
           Hikaru K{\sc awai}$^{1)}$\footnote
           {
e-mail address : kawaih@theory.kek.jp},\\
           Yoshihisa K{\sc itazawa}$^{2)}$\footnote
           {
e-mail address : kitazawa@th.phys.titech.ac.jp}{\sc and}
           Tsukasa T{\sc ada}$^{1)}$\footnote
           {
e-mail address : tada@theory.kek.jp}

        $^{1)}$ {\it High Energy Accelerator Research Organization (KEK),}\\
               {\it Tsukuba, Ibaraki 305, Japan} \\
        $^{2)}$ {\it Department of Physics,
Tokyo Institute of Technology,} \\
                 {\it Oh-okayama, Meguro-ku, Tokyo 152, Japan}\\
\end{center}

\vfill

\begin{abstract}
\noindent
We derive a long distance effective action for space-time coordinates
from a IIB matrix model.
It provides us an effective tool to study the structures of
space-time.
We prove the finiteness of the theory for finite $N$
to all orders of the perturbation theory.
Space-time is shown to be inseparable and its dimensionality
is dynamically determined.
The IIB matrix model contains a mechanism to ensure the vanishing cosmological
constant which does not rely on the manifest supersymmetry.
We discuss possible mechanisms to obtain
realistic dimensionality and gauge groups from the IIB matrix model.
\end{abstract}
\vfill
\end{titlepage}
\vfil\eject

\section{Introduction}
\setcounter{equation}{0}
A large $N$ reduced model has been proposed as a nonperturbative
formulation of type IIB superstring theory\cite{IKKT}\cite{FKKT}.
It is defined by the following action:
\beq
S  =  -{1\over g^2}Tr({1\over 4}[A_{\mu},A_{\nu}][A^{\mu},A^{\nu}]
+{1\over 2}\bar{\psi}\Gamma ^{\mu}[A_{\mu},\psi ]) ,
\label{action}
\eeq
here $\psi$ is a ten dimensional Majorana-Weyl spinor field, and
$A_{\mu}$ and $\psi$ are $N \times N$ Hermitian matrices.
It is formulated in a manifestly covariant way which we believe
is a definite advantage over the light-cone formulation\cite{BFSS}
to study the nonperturbative issues of superstring theory.
In fact we can in principle predict the dimensionality of space-time,
the gauge group and the matter contents by solving this model.
In this paper we report the results of our first such efforts toward that goal.

This action can be related to the Green-Schwarz action of superstring\cite{GS}
by using the semiclassical correspondence in the large $N$ limit:
\beqa
-i[\;,\;] &\rightarrow& {1\over N} \{\;,\;\}, \n
Tr &\rightarrow& N \int {d^2 \sigma }\sqrt{\hat{g}} .
\label{correspondence}
\eeqa
In fact eq.(\ref{action}) reduces to the Green-Schwarz action
in the Schild gauge\cite{Schild}:
\beq
S_{\rm Schild}=\int d^2\sigma [\sqrt{\hat{g}}\alpha(
\frac{1}{4}\{X^{\mu},X^{\nu}\}^2
-\frac{i}{2}\bar{\psi}\Gamma^{\mu}\{X^{\mu},\psi\})
+\beta \sqrt{\hat{g}}].
\label{SSchild}
\eeq
Through this correspondence, the eigenvalues of $A_{\mu}$ matrices are
identified with the space-time coordinates $X^{\mu}(\sigma )$.
The ${\cal N}=2$ supersymmetry manifests itself in  $S_{\rm Schild}$ as
\beqa
\delta^{(1)}\psi &=& -\frac{1}{2}
                      \sigma_{\mu\nu}\Gamma^{\mu\nu}\epsilon _1 ,\n
\delta^{(1)} X_{\mu} &=& i\bar{\epsilon _1}\Gamma_{\mu}\psi ,
\label{SSchildsym1}
\eeqa
and
\beqa
\delta^{(2)}\psi &=& \epsilon _2 ,\n
\delta^{(2)} X_{\mu} &=& 0 .
\label{SSchildsym2}
\eeqa
The ${\cal N}=2$ supersymmetry (\ref{SSchildsym1}) and (\ref{SSchildsym2}) is
directly translated into
the symmetry of $S$ as
\beqa
\delta^{(1)}\psi &=& \frac{i}{2}
                     [A_{\mu},A_{\nu}]\Gamma^{\mu\nu}\epsilon _1,\n
\delta^{(1)} A_{\mu} &=& i\bar{\epsilon _1}\Gamma_{\mu}\psi ,
\label{Ssym1}
\eeqa
and
\beqa
\delta^{(2)}\psi &=& \epsilon _2 ,\n
\delta^{(2)} A_{\mu} &=& 0.
\label{Ssym2}
\eeqa

The bosonic part of the action vanishes for the commuting matrices
$A_{\mu}^{ij}=x_{\mu}^i \delta ^{ij}$ where $i$ and $j$ are color indices.
These are the generic classical vacuum configurations.
Namely the distributions of the eigenvalues determine the extent and the
dimensionality
of space-time. Hence the structure of space-time is dynamically determined by
the theory.
As we will show in this paper, space-time exits as a single bunch
and no single eigenvalue can escape from the rest.
We may consider $n \times n$ submatrices
of our original $N \times N$ matrices
where $n$ is assumed to be still very large.
More precisely, 
we consider a block diagonal background where the first blocks are
$n$ dimensional and the rest is a diagonal vacuum configuration.
Such a background may represent a D-string (or D-objects) which occupies
a certain region of space-time.

Then the effective action which  describes the
$n\times n$ submatrices can be obtained by integrating the rest of the degrees
of freedom.
It has been shown that  the term proportional to $\beta$ in eq.(\ref{SSchild})
arises if the vacuum energy of order $N$ exists
in this model\cite{IKKT}.
In fact we will show in this paper that it is indeed the case.
The effective action may include all possible terms which are consistent with
the symmetry. The term proportional to $\beta$ is the lowest dimensional term
which is consistent with the symmetry and we expect that the higher
dimensional terms
are irrelevant for large $n$.
In this way we can understand why this model resembles the effective action
for D-objects\cite{Witten}. However we emphasize that we
regard the action (\ref{action}) as fundamental and it should be
distinguished from the
effective action for $N$ D-instantons.

It is straightforward to construct the configurations which represent
arbitrary numbers of D-objects in space-time in an analogous way.
Although they are constrained
to be inside the extension of space-time as we see in this paper,
otherwise they may be located anywhere.
As is shown in \cite{IKKT}, the long distance interactions of D-objects are
in accord with string theory.
Thus it must be clear that the IIB matrix model is definitely not the first
quantized
theory of a D-string but the full second quantized theory.
Arbitrary numbers of D-objects exist as local excitations of space-time
in this formulation which is  certainly not the case if we deal with the
effective theory of a fixed number of D-objects.

It has been also proposed that the Wilson loops are the creation and
annihilation
operators for strings.
We consider the following regularized Wilson loops\cite{FKKT}:
\beqa
w(C) & = & Tr[v(C)]  ,\n
v(C) &=&
\prod_{n=1}^M U_n ,\n
U_n &=&exp\{i\epsilon (k_{n}^{\mu}A_{\mu}
+\bar{\lambda }_{n}\psi )\} .
\label{Wilsonloop}
\eeqa
Here $k_n^{\mu}$ are momentum densities distributed along a loop $C$, and
we have also introduced the fermionic sources
$\lambda _{n}$ .
$\epsilon$ in the argument of the exponential
is a cutoff factor.
As it has been shown, $\ep$ can be regarded as a lattice
spacing of the worldsheet. We therefore call $\ep \rightarrow 0$ limit
the continuum limit.
This model has been further investigated through the loop equations which
govern the Wilson loops.
Such investigations have shown that the theory has no infrared divergences
and the string perturbation theory can be reproduced in the double
scaling limit.

It is possible to see the infrared finiteness of the theory in a more
direct way.
We may expand the fields around the vacuum configurations
$A_{\mu}^{ij}=x_{\mu}^i \delta ^{ij}$.
While the off-diagonal elements
of the matrices possess the nonvanishing propagators, the diagonal
components do not
possess the propagators and may be called the zeromodes.
We denote the bosonic zeromodes as $x_{\mu}^i$ and the fermionic zeromodes
as $\xi_{\alpha}^i$ where $i$ denotes the color indices.

The perturbative expansion around this background at first sight
appears to resemble that of the
original large $N$ reduced models\cite{RM}.
If so, the perturbative expansion around the
classical vacuum is identical to the ten dimensional super Yang-Mills theory
in the large $N$ limit.
The ultraviolet divergences of ten dimensional super Yang-Mills theory appear
as the infrared divergences in this model and hence the theory appears to
be ill-defined.
However the existence of the fermionic zeromodes turns out to modify
such a picture completely as will be shown in this paper.

The possibly dangerous configurations are such that the eigenvalues $x_{\mu}^i$
are widely separated compared to the scale set by $g^2$.
It is clear that the mass scale of the
off-diagonal elements and that of the zeromodes are also
widely separated in such a region. Therefore we can adopt a
Born-Oppenheimer type
approximation and we can integrate the off-diagonal elements first.
Here the expansion parameter is $g^2$  over the fourth power of the
average distance between the eigenvalues .
Therefore the one loop approximation gives the accurate result for
widely separated eigenvalues.
In this way we can obtain the effective Lagrangian for the zeromodes.

We can subsequently perform the
integration over the fermionic zeromodes.
The resulting effective action for the bosonic zeromodes turns out
to be completely infrared finite for finite $N$.
Since the theory is manifestly ultraviolet finite, we have established
the remarkable fact that this model is a finite theory which is free
from both short distance and long distance divergences.

The effective action provides us a useful tool to study the structures
of space-time. In section three, we study it
to determine the distributions of space-time coordinates.
We believe that we have entered a new paradigm in string theory where
we are beginning to determine the structure of space-time dynamically.
The readily recognizable structure which emerges
from our investigation is in fact a four dimensional object
albeit it is a fractal.
We find that space-time points form branched polymers
in ten dimensions within the simplest approximation.
Obviously, we need some mechanism to flatten it into four dimensions
to describe our space-time. Although our investigation to find it
in the IIB matrix model is still in progress, we explicitly
propose possible mechanisms.

We remark that related models to ours have been proposed and studied in
\cite{periwal}$\sim$\cite{oda}.
We also note a deep connection between our approach and the noncommutative
geometry
\cite{Connes}.
This paper consists of four sections.
In the first introductory section, we have reviewed the conceptual
framework of our approach.
In section two, we derive the long distance effective action for the
zeromodes(super coordinates).
In section three, we study the effective action to understand the possible
structure of space-time.
Section four is devoted to the conclusions and the discussions.
One of the crucial issues in our approach is how to take the large $N$ limit.
We address this question in the concluding section.

\setcounter{equation}{0}
\section{Effective theory for diagonal elements}\label{sec:effdia}
In this section, we discuss an effective theory for the diagonal
elements of the IIB matrix model.
As we have explained in the introduction, the theory classically
possesses the huge moduli and the diagonal elements of $A_{\mu}$ may
assume any values. However this classical degeneracy is lifted
quantum mechanically and there remains no moduli in the theory as
we shall show in this section.
Let us consider the expansion around the most generic classical
moduli where the gauge group $SU(N)$ is completely broken down to
$U(1)^{N-1}$.
Then the diagonal elements of $A_{\mu}$ and $\psi$ appear
as the zeromodes while the off-diagonal elements become massive.
So we may integrate out the massive modes first and obtain the effective
action for the diagonal elements.

We thus decompose $A_{\mu}$ into diagonal part $X_{\mu}$
and off-diagonal part $\tilde{A}_{\mu}$.
We also decompose $\psi$ into $\xi$ and $\tilde{\psi}$:
\beqa
A_{\mu}&=& X_{\mu} + \tilde{A}_{\mu}; \;\;
X_{\mu}= \left( \begin{array}{llll}
                 x_{\mu}^1 &&& \\
                 &x_{\mu}^2 && \\
                 && \ddots &   \\
                  &&& x_{\mu}^N
         \end{array} \right), \n
\psi &=& \xi + \tilde{\psi}; \;\;
\xi = \left( \begin{array}{llll}
                 \xi^1 &&& \\
                 &\xi^2 && \\
                 && \ddots &   \\
                  &&& \xi^N
         \end{array} \right),
\eeqa
where $x^i_\mu$ and $\xi^i_\alpha$ satisfy the constraints 
$\sum_{i=1}^N x^i_\mu=0$ and $\sum_{i=1}^N \xi^i_\alpha =0$, respectively,
since the gauge group is $SU(N)$.
We then integrate out the off-diagonal parts $\tilde{ A_{\mu}}$
and $\tilde{\psi}$ and obtain the
effective action for supercoordinates of space-time $S_{\rm eff}[X,\xi]$.
The effective action for space-time coordinates $S_{\rm eff}[X]$
can be obtained by further integrating out $\xi$:
\beqa
\int dAd\psi e^{-S[A,\psi]}
&=& \int dXd\xi e^{-S_{\rm eff}[X,\xi]} \nonumber \\
&=& \int dX e^{-S_{\rm eff}[X]},
\eeqa
where $dX$ and $d\xi$ stand for $\prod_{i=1}^{N-1} dx^i_\mu$
and $\prod_{i=1}^{N-1} d\xi^i_\alpha$, respectively.

\subsection{Perturbative evaluation of $S_{\rm eff}[X,\xi]$}
\label{sec:intoffdiag}

In this subsection, we integrate out the
off-diagonal elements $\tilde{A}_{\mu}$ and $\tilde \psi$
by perturbative expansion in $g^2$.
As we will see below,
this expansion is valid
when all of the diagonal elements are widely separated from one another.
Since $X_{\mu}$ and $\xi$ do not have propagators, we treat them as
collective coordinates.

The original action (\ref{action}) can be expanded as follows,
\begin{eqnarray}
S&=&S_2 + S_{\rm int} \label{eq:decompose S}\\
&S_2 &= \frac{1}{2g^2} Tr\bigl(  - [X_\mu,\tilde A_\nu][X^\mu,\tilde
A^\nu]
+[X_\mu, \tilde A_\nu][X^\nu, \tilde A^\mu] \nonumber \\
&&\qquad \qquad -\bar{\tilde \psi} \Gamma^\mu[X_\mu, \tilde \psi]-
 [ \bar{\xi}, \tilde A_\mu] \Gamma^\mu \tilde \psi -
 \bar{\tilde \psi}\Gamma^\mu[\tilde A_\mu,\xi] \bigr),  \label{eq:S2}\\
& S_{\rm int}&= \frac{1}{2g^2} Tr\bigl(  - 2 [X_\mu,\tilde A_\nu][\tilde
A^\mu,\tilde A^\nu]
-\frac12[\tilde A_\mu, \tilde A_\nu][\tilde A^\mu, \tilde A^\nu] \nonumber \\
&& \quad  \quad \qquad - \bar{\tilde \psi}\Gamma^\mu[\tilde A_\mu,\tilde\psi]
\bigr). \label{eq:Sint}
\end{eqnarray}
We can observe that $\xi$ has no propagator
since the quadratic term of $\xi$ is absent.
As for the gauge fixing, we adopt the following covariant gauge:
\footnote{Actually there is subtlety for the ghost zeromode
in the covariant gauge.
However the same one loop effective action can be
obtained without such subtlety in the light cone gauge 
where $A_{+}$ is diagonalized.}
\beqa
S_{\rm g.f.}&=& -\frac{1}{2g^2} Tr ([X_\mu, A^\mu]^2),
\nonumber \\
S_{\rm F.P.} &=& - \frac{1}{g^2}Tr ([X_\mu,b][A^\mu,c]),
 \eeqa
where $b$ and $c$ are the Faddeev-Popov ghost fields.

In terms of the components, $S_2+S_{\rm g.f.}$ can be written as
\begin{eqnarray}
S_2+S_{\rm g.f.}
&=&\frac{1}{2g^2} \sum_{i<j}
\bigl( (x_\nu^i-x_\nu^j)^2 {{\tilde A}^{ij}_\mu}{}^*
{{\tilde A}^{ij}}{}^\mu
- \bar{\tilde \psi}^{ji}\Gamma^\mu (x^i_\mu-x^j_\mu) \tilde \psi^{ij}\n
&&+ (\bar{\xi}^i-\bar{\xi}^j) \Gamma^\mu \tilde \psi^{ij}
{\tilde{A_\mu^{ij}}}^*
+\bar{\tilde \psi}^{ji} \Gamma^\mu ({\xi}^i-{\xi}^j){\tilde{A_\mu^{ij}}}
\bigr).
\label{eq:S2comp}
\end{eqnarray}
The first and the second terms are the kinetic
terms for $\tilde{A}$ and $\tilde{\psi}$ respectively,
while the last two terms are
$\tilde{A}\tilde{\psi}\xi$ vertices.
The basic building blocks of the Feynman rules are:
\beq
<\tilde A_\mu^{ij} \tilde A_\nu^{*}{}^{ij}>\equiv
\begin{picture}(100,20)(0,10)
\Photon(50,15)(100,15){2}{4}
\Photon(100,11)(50,11){2}{4}
\Text(75,26)[]{$i$}
\Text(75,0)[]{$j$}
\Text(40,12)[]{$ \mu$}\Text(110,12)[]{$ \nu$}
\end{picture}
\qquad=\frac{\eta_{\mu\nu}}{(x^i-x^j)^2}
\label{eq:bosonprop}
\eeq
\beq
<\tilde \psi^{ij} \bar{\tilde {\psi^{ij}}}>\equiv
\begin{picture}(100,20)(10,10)
\ArrowLine(50,15)(100,15)
\ArrowLine(100,11)(50,11)
\Text(75,26)[]{$i$}
\Text(75,0)[]{$j$}
\end{picture}
=-\frac{1}{(x^i-x^j)_\mu \Gamma^\mu}
\label{eq:fermionprop}
\eeq
\beq
\tilde{A}\tilde{\psi}\xi\; {\rm vertex}\equiv
\begin{picture}(100,40)(10,5)
\ArrowLine(50,15)(72,15)
\ArrowLine(72,11)(50,11)
\Text(60,26)[]{$i$}
\Text(60,0)[]{$j$}
\Photon(77,15)(100,15){2}{2}
\Photon(100,11)(77,11){2}{2}
\Text(90,26)[]{$i$}
\Text(90,0)[]{$j$}
\GCirc(75,13){4}{1}
\Text(110,12)[]{$ \mu$}
\Line(75,10)(75,0)
\end{picture}
\qquad=\Gamma^\mu(\xi^i-\xi^j)
\label{eq:xivertex}
\eeq
As $ \tilde A$ and $ \tilde \psi$
are matrices, the propagators are denoted by double lines with indices $ i$
and $ j$.

The interaction terms in \eq{eq:Sint} are denoted by the following vertices:
\begin{equation}
\hbox{boson three-point vertex}\equiv
\begin{picture}(100,50)(-20,35)
\Photon(15,60)(35,43){2}{3}
\Photon(35,43)(55,60){2}{3}
\Photon(59,55)(37,38){2}{3}
\Photon(32,38)(12,55){2}{3}
\Photon(37,38)(37,15){2}{3}
\Photon(32,15)(32,38){2}{3}
\Text(35,60)[]{$i$}
\Text(50,30)[]{$j$}
\Text(20,30)[]{$k$}
\Text(35,7)[]{$\mu$}
\Text(8,62)[]{$\nu$}
\Text(62,62)[]{$\lambda$}
\end{picture}
\end{equation}
\begin{equation}
\hbox{boson four-point vertex} \equiv
\begin{picture}(100,50)(-20,35)
\Photon(15,60)(35,43){2}{3}
\Photon(35,43)(55,60){2}{3}\Photon(59,55)(40,40){2}{3}
\Photon(40,40)(60,23){2}{3}\Photon(56,18)(35,37){2}{3}
\Photon(35,37)(14,18){2}{3}
\Photon(12,23)(30,40){2}{3}\Photon(30,40)(12,55){2}{3}
\Text(35,60)[]{$i$}
\Text(60,40)[]{$j$}
\Text(35,20)[]{$k$}
\Text(10,40)[]{$l$}
\Text(62,62)[]{$\mu$}
\Text(62,16)[]{$\nu$}
\Text(8,16)[]{$\lambda$}
\Text(8,62)[]{$\rho$}
\end{picture}
\end{equation}
\begin{equation}
\hbox{boson fermion vertex}\equiv
\begin{picture}(100,70)(-20,0)
\ArrowLine(0,30)(60,30)
\ArrowLine(60,25)(33,25)
\ArrowLine(27,25)(0,25)
\Photon(33,25)(33,5){2}{3}
\Photon(27,5)(27,25){2}{3}
\Text(30,38)[]{$i$}
\Text(45,15)[]{$j$}
\Text(15,15)[]{$k$}
\Text(30,-3)[]{$\mu$}
\end{picture}
\end{equation}

One can see from the above Feynman rules that the expansion parameter 
after the $\xi$ integration is
$g^2$ over the fourth power of the average distances
between the space-time coordinates.
Therefore, the perturbative expansion is valid when all of the
diagonal elements are widely separated from one another.
In particular, one-loop approximation is valid in these regions.

A novel feature of the above Feynman rules compared with those
of the gauge theory in the large $N$ limit is the
presence of $\xi$ insertion vertices (\ref{eq:xivertex}).
If we were to set $ \xi =0 $,
the contributions from $\tilde A$ and $\tilde \psi$
along from the ghosts cancel at the one loop level:
\beq
\int d \tilde A d\tilde \psi db dc \;
e^{ -(S_2 +S_{\rm g.f.}+S_{\rm F.P.})}
=\prod_{i<j}(x^i-x^j)^{2(-10+2+8)} =1. \label{eq:xizero}
\eeq
Thus the diagonal elements would take any value and the classical moduli
would remain at least perturbatively.
The theory is identical to the large $N$ limit of the super Yang-Mills
theory if we identify $X$ with momenta\cite{RM}.
Therefore the major difference between the IIB matrix model and
super Yang-Mills theory is the presence of dynamical zeromodes
$X$ and $\xi$ in the former.
Although the zeromodes are order $N$ quantity,
we cannot ignore them even in the large $N$ limit as we shall
show below.

Eq. (\ref{eq:S2comp}) shows that the one loop effective action
can be expressed by a super determinant.
It turns out to be most useful to integrate out $\tilde{\psi}$
first.
Such an integration gives rise to the following quadratic term
of $\tilde{A}_{\mu}$:
\begin{equation}
\frac{1}{g^2} \sum_{i<j} (\bar{\xi^i} -\bar{\xi^j}) \Gamma^{\mu\alpha\nu}
(\xi^i-\xi^j) \frac{(x^i_\alpha-x^j_\alpha)}{(x^i-x^j)^2}
\tilde{A}^{ij*}_\mu \tilde{A}^{ij}_\nu,
\end{equation}
where $\Gamma ^{\mu\alpha\nu}={1\over 3!}\Gamma ^{[\mu}\Gamma ^{\alpha}
\Gamma ^{\nu ]}$.
Here the indices within the bracket $[\;]$ are totally antisymmetrized.
In this way we obtain the following one-loop effective
action for the zeromodes:
\begin{eqnarray}
 \int d \tilde A d\tilde \psi d b dc \;
e^{ - (S_2 +S_{\rm g.f.}+S_{\rm F.P.})}
&=& \prod_{i<j} {\rm det}_{\mu\nu}
\bigl( \eta^{\mu\nu}+S^{\mu\nu}_{(ij)}\bigr)^{-1} \nonumber\\
&\equiv&
e^{-S_{\rm eff}^{\rm 1-loop} [X, \xi]} ,
\end{eqnarray}
where
\begin{equation}
S^{\mu\nu}_{(ij)}=(\bar{\xi^i} -\bar{\xi^j}) \Gamma^{\mu\alpha\nu}
(\xi^i-\xi^j) \frac{(x^i_\alpha-x^j_\alpha)}{(x^i-x^j)^4}. \label{eq:defS}
\end{equation}

The effective action can be expanded as
\begin{eqnarray}
S_{\rm eff}^{\rm 1-loop} [X, \xi]
&=& \sum_{i<j} tr \ln(\eta^{\mu\nu}+S^{\mu\nu}_{(ij)}) \nonumber \\
&=& - \sum_{i<j} tr
\bigl(\frac{S^2_{(ij)}}{2}+\frac{S^4_{(ij)}}{4}+\frac{S^6_{(ij)}}{6}+\frac{S
^8_
{(ij)}}{8} \bigr).
\label{eq:seff1}
\end{eqnarray}
Here the symbol $tr$ in the lower case stands for the trace for Lorentz indices.
Note that
the expansion terminates with the eighth power of $S^{\mu\nu}_{(ij)}$,
since $\xi$ has 16 spinor components and $S^{\mu\nu}_{(ij)}$
is bilinear in $\xi$.
Note also
that  only even powered terms remain,
since $S^{\mu\nu}_{(ij)}$ is anti-symmetric under the interchange of
$ \mu$ and $ \nu$.
Furthermore the quadratic and sextic terms vanish identically
as is shown in the appendix \ref{sec:fierz}.
Therefore we obtain
\begin{equation}
S_{\rm eff}^{\rm 1-loop} [X, \xi]=- \sum_{i<j} tr
\bigl(\frac{S^4_{(ij)}}{4}
+\frac{S^8_{(ij)}}{8} \bigr) .\label{eq:seff}
\end{equation}
We remark that eq. (\ref{eq:seff}) is gauge independent.
It is because the longitudinal part of the bosonic propagator
which is sensitive to the gauge parameter
has no contribution since
\begin{equation}
(x^i_{\mu}-x^j_{\mu}) S^{\mu\nu}_{(ij)}
=(\bar{\xi^i} -\bar{\xi^j}) \Gamma^{\mu\alpha\nu}
(\xi^i-\xi^j) \frac{(x^i_{\mu}-x^j_{\mu})(x^i_\alpha-x^j_\alpha)}{(x^i-x^j)^4}
=0 .
\end{equation}

It is also possible to construct similar reduced models for super Yang
Mills theory in $D$=3,4 and 6 by the large $N$ reduction procedure.
Then the calculation we have explained in this section
is also applicable to these cases and we find
\begin{equation}
S_{\rm eff}^{\rm 1-loop} [X, \xi]=\cases{
0 & for  $D$=3 \cr
- \sum_{i<j} tr \bigl(\frac{S^2_{(ij)}}{2}
 \bigr) & for $D$=4 \cr
- \sum_{i<j} tr \bigl(\frac{S^2_{(ij)}}{2}
+\frac{S^4_{(ij)}}{4} \bigr) & for $D$=6
} .
\label{grsym}
\end{equation}
Before we proceed to perform $\xi$ integration and to
obtain the effective action for $X$,
we discuss $ {\cal N}=2$ supersymmetry of the effective action 
in the next subsection.

\subsection{${\cal N}=2$ SUSY}\label{sec:susy}
In this subsection,
we show  that the effective action (\ref{eq:seff}) has the following $ {\cal
N}=2$ supersymmetry:
\begin{equation}
\cases{
\delta^{(1)} x^i_\mu &  $= i \bar{\epsilon_1}\Gamma_\mu \xi^i $\cr
\delta^{(1)} \xi ^i& =0
}, \label{eq:susya}
\end{equation}
\begin{equation}
\cases{
\delta^{(2)} x^i_\mu & = 0\cr
\delta^{(2)} \xi ^i& $=\epsilon_2$
}. \label{eq:susyb}
\end{equation}
These symmetries are the remnants of the symmetry in the original theory:
\beqa
&&\cases{
\delta^{(1)} A^i_\mu &  $= i \bar{\epsilon_1}\Gamma_\mu \psi $\cr
\delta^{(1)} \psi & $=\frac{i}{2}\Gamma^{\mu\nu} [A^\mu, A^\nu] {\epsilon_1}$
},  \label{eq:susyao}\\
&&\cases{
\delta^{(2)} A^i_\mu & = 0\cr
\delta^{(2)} \psi & $=\epsilon_2$
}.
\label{eq:susybo}
\eeqa
First let us decompose the variables
into the diagonal and off-diagonal
parts in the original transformations (\ref{eq:susyao}) and (\ref{eq:susybo}).
We may consider $\tilde{A_{\mu}}$ and $\tilde{\psi}$
as the quantities of order $\hbar$.
Then one can expand these transformations in terms of $\hbar$.
However the action is invariant under these transformations
at each order of $\hbar$.
Thus $S_2$ (\ref{eq:S2}) is invariant under the
transformations which are linear in $\tilde{A_{\mu}}$ and $\tilde{\psi}$,
\begin{equation}
\cases{
\delta^{(1)} x^i_\mu &  $= i \bar{\epsilon_1}\Gamma_\mu \xi^i $\cr
\delta^{(1)} \tilde A^{ij}_\mu &  $= i \bar{\epsilon_1}\Gamma_\mu \tilde \psi^{ij}
$\cr
\delta^{(1)} \xi^i & $=0 $ \cr
\delta^{(1)} \tilde\psi^{ij} &$ = i (x^i-x^j)_\mu \tilde A^{ij}_\nu
\Gamma^{\mu\nu}\epsilon_1$
}, \label{eq:susyfe}
\cases{
\delta^{(2)} x^i_\mu &  $= 0$\cr
\delta^{(2)} \tilde A^{ij}_\mu &  $= 0$\cr
\delta^{(2)} \xi^i &$  =\epsilon_2$ \cr
\delta^{(2)} \tilde\psi^{ij} &$= 0$
}.\end{equation}
After the integration over $ \tilde A$ and $ \tilde \psi$, the remaining
effective action (\ref{eq:seff})
shall have the symmetries (\ref{eq:susya}) and (\ref{eq:susyb}).

One can also show the invariance of (\ref{eq:seff}) under the transformations
(\ref{eq:susya}) and (\ref{eq:susyb}) through explicit calculations.
Since $S^{\mu\nu}_{(ij)}$ contains $ \xi^i$ only through the
combination as $ \xi^i-\xi^j$,
one can see that the invariance under 
(\ref{eq:susyb}) is satisfied rather trivially.
As for (\ref{eq:susya}), it introduces additional $ \xi^i$.
$ tr (S_{(ij)})^8$ is invariant under (\ref{eq:susya}),
since it already contains 16 $ \xi$'s.
Some calculations are required to exhibit the invariance of the term 
$tr (S_{(ij)})^4$:
\beq
tr \delta^{(1)} S_{(ij)} S^3_{(ij)}
=i \frac{1}{(x^4)^4}x_\lambda x_\rho x_\sigma
U_\alpha^{\;\;\alpha \lambda \rho \sigma}
- 4i \frac{1}{(x^4)^4}\frac{1}{x^2}x_\mu x_\nu
x_\lambda x_\rho x_\sigma U^{\mu \nu \lambda \rho \sigma},
\eeq
where $U^{\mu \nu \lambda \rho \sigma}=
 (\bar{\epsilon_1}\Gamma^\mu \xi)
(\bar{\xi}\Gamma_{\alpha}^{\;\;\nu\beta}\xi)
(\bar{\xi}\Gamma_{\beta}^{\;\;\lambda\gamma} \xi)
(\bar{\xi}\Gamma_{\gamma}^{\;\;\rho\delta} \xi)
(\bar{\xi}\Gamma_{\delta }^{\;\;\sigma\alpha} \xi)
$
is a tensor which is totally symmetric and traceless with respect to
the indices  $\nu\lambda\rho\sigma$
as is explained in the appendix A.  Here $ x_\mu$ and $\xi$
denote $ (x_\mu^i-x_\mu^j)$ and $ \xi^i-\xi^j$ respectively.
We may further decompose  $U^{\mu \nu \lambda \rho \sigma}$
into the irreducible components as
\begin{equation}
U^{\mu \nu \lambda \rho \sigma}_{\rm{sym.traceless}}=\hbox{symmetrization of
}U^{\mu \nu \lambda \rho \sigma}-\frac{1}{40}(g^{\mu\nu}U_\alpha^{\;\;\alpha
\lambda \rho \sigma}+{\rm 9\; terms}).
\end{equation}
And we can show that $U^{\mu \nu \lambda \rho
\sigma}_{\rm {sym.traceless}}=0$ by decomposing SO(10) $\supset$ SO(2) $
\times$SO(8) where $ \xi=(s_a, c_{\dot{a}}), \epsilon={(b_a,
t_{\dot{a}})}$.
It is then sufficient to demonstrate that
$U^{+++++}_{\rm{sym.traceless}}=s^9 b=0$.
Therefore the following cancellation implies the invariance of
$ tr (S_{(ij)})^4$:
\begin{equation}
tr \delta^{(1)} S_{(ij)} S^3_{(ij)}=i \frac{1}{(x^4)^4}(1-4\frac{10}{40})x_\lambda
x_\rho x_\sigma U_\alpha^{\;\;\alpha\lambda \rho \sigma}=0.
\end{equation}
This completes the proof of the invariance of (\ref{eq:seff}) under
(\ref{eq:susya})
and (\ref{eq:susyb}).

\subsection{Dynamics of X at long distances}\label{sec:intxi}

In this subsection, we study the integration procedure of $\xi$
to obtain the effective theory for the space-time coordinates $X$:
\beq
\int dX d\xi \: e^{ -S_{\rm eff}[X,\xi]}
= \int dX\: e^{-S_{\rm eff}[X]}.
\label{eq:intxi}
\eeq
We perform $\xi$ integrations explicitly at the one loop
level in this subsection and explain that it can be represented
graphically in terms of trees which connects the space-time points.
Although the vacuum energy of order $N^2$ has been canceled between the
bosonic and fermionic degrees of freedom,
it does not vanish at order $N$ in the IIB matrix model since
$S_{\rm eff}[X]$ is shown to be nontrivial.
Through the perturbative evaluation of $S_{\rm eff}[X]$ to all orders,
we prove the finiteness of the $X$ integration in \eq{eq:intxi} for finite $N$.

\subsubsection{One loop evaluation of $S_{\rm eff}[X]$}

We substitute the one loop effective action $S_{\rm eff}^{\rm 1-loop}[X,\xi]$
of eq. (\ref{eq:seff}) into \eq{eq:intxi}
and consider what kind of terms survive after $\xi$ integrations:
\beqa
\int dXd\xi e^{-S^{\rm 1-loop}_{\rm eff}[x,\xi]}
&=& \int dXd\xi \prod_{i<j}[1+\frac{tr(S_{(ij)}^4)}{4}
                             +(\frac{1}{2}(\frac{tr(S_{(ij)}^4)}{4})^2
                               +\frac{tr(S_{(ij)}^8)}{8})]
\label{eq:mulprod}
\eeqa
Here the products  are taken over all possible different
pairs of color indices $(ij)$.
When we expand the multi-products, we select one of the three different
factors
$1$ or $tr(S^4_{(ij)})/4$  or  $(tr(S^8_{(ij)})/8+ (tr(S^4_{(ij)}))^2/32 )$ for 
each pair of $(ij)$. Since the last two  factors are functions of $(x^i-x^j)$,
they can be visualized by bonds that  connect the ``space-time
points'' $x^i$ and $x^j$. 
More precisely,  in order to remind us that
the factors $tr(S^4_{(ij)})/4$
and  $(tr(S^8_{(ij)})/8+ (tr(S^4_{(ij)}))^2/32 )$ 
contain 8 and 16 spinor components of $\xi^i_\alpha-\xi^j_\alpha$,
 we draw 8 or 16 bonds between $x^i$ and $x^j$
depending on whether we take 
$tr(S^4_{(ij)})/4$ or  $(tr(S^8_{(ij)})/8+ (tr(S^4_{(ij)}))^2/32 )$.
That is, each bond corresponds to each component of a spinor
$\xi^i_\alpha-\xi^j_\alpha$. 
We call these sets of 8 and 16 bonds ``8-fold bond''
and ``16-fold bond'' respectively.
In this way  we can associate each term in the 
 expansion of multi-products in \eq{eq:mulprod}  with a graph 
connecting the space-time points by 8-fold bonds  or 16-fold  bonds.
We don't assign any  bond to the factor 1.
Therefore the multi-products in \eq{eq:mulprod} can be replaced by a summation
over all possible graphs.

Let us consider spinor $\xi^i_\alpha$ component by component 
in order to discuss
what kind of graphs survive
after $\xi$ integrations.
Out of the 16 spinor components of $\xi^i_{\alpha}$, we may focus on a particular
spinor component
such as the first component $\xi^i_1$.
We rewrite \eq{eq:mulprod} as
\beq
\int dX \int \prod_{i=1}^{N-1} d\xi^i_2 \cdots d\xi^i_{16}
             \prod_{i=1}^{N-1} d\xi^i_1
\prod_{i<j}(C_0+C_1 \cdot(\xi_1^i-\xi_1^j)).
\eeq
Here $C_0$ and $C_1$ are functions of $x^i_{\mu}-x^j_{\mu}$
and the other spinor components, 
$\xi^i_2-\xi^j_2, \cdots, \xi^i_{16}-\xi^j_{16}$. 
We can then expand the multi-products regarding the factor
$C_1 \cdot (\xi_1^i-\xi_1^j)$
as a single bond.
Then it is clear from the following considerations
that the integrations of $\xi ^i_1$ for
all color indices $i$ generate subgraphs 
associated with the first component of spinors,
which connect all the $N$ points without a loop.
Such type of graphs are called maximal trees.\\
i)Since $\xi ^i_1$ has $N-1$  independent color components,
the subgraphs having $N-1$ bonds remain after $\xi ^i_1$ integration.\\
ii)If there is a loop in the subgraph, the contribution vanishes
since a product  of delta functions of grassmann variables on the loop
vanishes:
\beq
\delta(\xi_1^{i_1}-\xi_1^{i_2})
\delta(\xi_1^{i_2}-\xi_1^{i_3})
\cdots
\delta(\xi_1^{i_{(k-1)}}-\xi_1^{i_k})
\delta(\xi_1^{i_k}-\xi_1^{i_1})=0.
\eeq
We also note that all maximal trees contribute equally
as we can see by performing $\xi_1^i$ integrations form the end points
of the maximal trees.

Therefore the whole integration of the fermionic degrees of freedom
generates graphs which are superpositions of 
16 maximal trees.
However these 16 maximal trees are not independent.
At each bond they are constrained to be bunched into
the set of 8 or 16.
From these considerations, we find a graphical representation of
the one loop effective action:
\beqa
\int dXd\xi e^{-S^{\rm 1-loop}_{\rm eff}[x,\xi]}
&=& \int dXd\xi \sum_{G:{\rm graph}}\;\; \prod_{(ij):{\rm bond\; of \;} G}\n
&&               [(\frac{tr(S_{(ij)}^4)}{4})\;\; {\rm or}\;\;
                (\frac{1}{2}(\frac{tr(S_{(ij)}^4)}{4})^2
                +\frac{tr(S_{(ij)}^8)}{8})].
\label{eq:graph}
\eeqa
Here we sum over all possible graphs 
consisting of 8-fold and 16-fold bonds
which can be expressed as superpositions of 16 maximal trees.
For each bond $(ij)$ of $G$ we assign the first or the second factor 
depending whether it is 8-fold or 16-fold bond.

Since each 16-fold bond contains 16 spinors,
$(\xi^i-\xi^j)$ and $(x^i-x^j)$ must form Lorentz singlets by themselves,
\beq
(\frac{1}{2}(\frac{tr(S_{(ij)}^4)}{4})^2+\frac{tr(S_{(ij)}^8)}{8})
\sim \delta^{(16)}(\xi^i-\xi^j)\;\frac{1}{(x^i-x^j)^{24}}.
\label{eq:dbt}
\eeq
In the 8-fold bond, however, $(\xi^i-\xi^j)$ and $(x^i-x^j)$
couple as
\beq
tr(S_{(ij)}^4)=T^{\mu\nu\lambda\rho}
(x^i-x^j)_{\mu}(x^i-x^j)_{\nu}(x^i-x^j)_{\lambda}(x^i-x^j)_{\rho}
/(x^i-x^j)^{16}.
\label{eq:sbt}
\eeq
Here
\beqa
T^{\mu\nu\lambda\rho}&=&
[(\bar{\xi^i}-\bar{\xi^j})\Gamma^{\mu\alpha}_{\;\;\;\;\beta}(\xi^i-\xi^j)]
[(\bar{\xi^i}-\bar{\xi^j})\Gamma^{\nu\beta}_{\;\;\;\;\gamma}(\xi^i-\xi^j)]\n &&
[(\bar{\xi^i}-\bar{\xi^j})\Gamma^{\lambda\gamma}_{\;\;\;\;\delta}(\xi^i-\xi^j)]
[(\bar{\xi^i}-\bar{\xi^j})\Gamma^{\rho\delta}_{\;\;\;\;\alpha}(\xi^i-\xi^j)]
\label{tmn}
\eeqa
is a totally
symmetric traceless tensor as is shown in the appendix \ref{sec:fierz}.

If there were only 16-fold bonds,
considerable simplifications take place
since the graphs are reduced to maximal trees with 16-fold degeneracy
and the interactions between the points
only depend on their distances.
We may cite the four dimensional model in eq. (\ref{grsym})
as a simple model which shares such a property.
The effective action of the four dimensional model
can be represented as follows:
\beqa
\int dXd\xi e^{-S^{\rm 1-loop}_{\rm eff}[x,\xi]}
&=& \int dXd\xi \prod_{i<j}[1+\frac{tr(S_{(ij)}^2)}{2}]\nonumber\\
&=& \int dXd\xi \sum_{G:{\rm graph}}\; \prod_{(ij):{\rm bond\;of\;}G}
                                     \frac{tr(S_{(ij)}^2)}{2},
\label{eq:graph4dim}
\eeqa
where we sum over all graphs whose bonds form maximal trees.
Since $\xi$ has four spinor components in four dimensions,
\beq
tr(S_{(ij)}^2)
\sim \delta^{(4)}(\xi^i-\xi^j)\;\frac{1}{(x^i-x^j)^{6}}.
\label{eq:xi4x6}
\eeq
Therefore, in the four-dimensional model,
the distribution of $X$ becomes ``branched polymer'' type:
\beq
\int dX e^{-S_{\rm eff}^{\rm 1-loop}[X]} =
\int dX \sum_{G: {\rm maximal\; tree}}\; \prod_{(ij):{\rm bond\;of\;}G}
\frac{1}{(x^i-x^j)^6}.
\label{eq:bp}
\eeq

Note that all points are connected by the bonds, and
each $x^{ij}$ integration can be performed independently and 
converges for large $x^{ij}$ on each bond
where $x^{ij}=x^i-x^j$.
Thus this system is infrared convergent.
Although the integrations seem to be divergent for short distances,
it is due to the failure of the one loop approximation since
the theory itself is manifestly finite at short distances.
We will consider the short distance behavior of the theory
in subsection
\ref{sec:short distance}.
As we see in the appendix \ref{app:BP},
the dynamics of branched polymer is well known
and its Hausdorff dimension is four.
So this model constitutes an example of models for dynamical
generation of space-time
which predicts four dimensional space-time.
As is shown in the subsequent discussions,
the IIB matrix model is much more complicated.
Nevertheless
we expect that the structure of space-time is also determined dynamically
in the IIB matrix model.

\subsubsection{Infrared convergence}\label{sec:IRconv}

In what follows we show that the $X$ integral is convergent
in the infrared region for finite $N$ to all orders of
perturbation theory:
\beq
\int dX e^{-S_{\rm eff}[X]} < \infty \qquad({\rm infrared}).
\label{eq:IRconv}
\eeq
It shows that all points are gathered as a single bunch
and hence space-time is inseparable.
Indeed we can show that
{\it the multiple integral, {\rm \eq{eq:IRconv}}, is absolutely convergent}.

We use the following theorem to prove the above statement:
{\it An $M$-dimensional multiple integral is absolutely convergent when
the superficial degrees of divergences of sub-integrals
on any
$m$-dimensional hyper-planes within ${\bf R}^M$ are negative.}
Let us apply the identical linear transformations
$T$ to the variables $X$ and $\xi$ as
\beq
\left(\begin{array}{c}
y^1\\
\vdots\\
y^{N-1}
\end{array}\right)
=T
\left(\begin{array}{c}
x^1\\
\vdots\\
x^{N-1}
\end{array}\right),\qquad
\left(\begin{array}{c}
\eta^1\\
\vdots\\
\eta^{N-1}
\end{array}\right)
=T
\left(\begin{array}{c}
\xi^1\\
\vdots\\
\xi^{N-1}
\end{array}\right),
\eeq
and consider the case where the values of $y^1 \cdots y^m$ become large
while $y^{m+1} \cdots y^{N-1}$ are fixed.
We observe from the Feynman rules in subsection \ref{sec:intoffdiag}
that we can associate the factor $(x^i-x^j)^{-{3\over 2}}$
with every $\xi^i-\xi^j$ by absorbing half of the powers of the
bosonic and the fermionic propagators which are connected to $\xi^i-\xi^j$
in \eq{eq:xivertex}.
Although $X$ dependencies of the diagrams are not entirely  absorbed 
by this procedure
at higher orders, the remaining factors make the infrared convergence
properties even better.
We explain this point by concrete examples at the end of this subsection.

We can reexpress $(x^i-x^j)^{-{3\over 2}}$ and $\xi^i-\xi^j$
in terms of $y$ and $\eta$ through the
identical linear transformations as $ (\sum _k C^{ij}_k y^k)^{-{3\over 2}}$ and
$ \sum _k C^{ij}_k \eta^k$ respectively.
The perturbative expansion of the effective action
consists of many terms which contain fermionic variables through the
combinations of $\xi^i-\xi^j$.
Let us consider the contribution from a
term which contains $\eta^1_1$.
If it is associated with the bond $(ij)$,
it means that $C^{ij}_1\ne0$ and we obtain the factor
$ ( C^{ij}_1 y^1 + \cdots)^{-{3\over 2}}$ 
after the integration of $\eta^1_1$.
We obtain the analogous factors from the integrations over 
all spinor components of fermionic variables
up to $\eta^m$.
Thus we can conclude that
\beq
\mid e^{-S_{\rm eff}[y]} \mid < 
\frac{1}{\prod_{i=1}^m \prod_{\alpha=1}^{16} P^i_\alpha} ,
\eeq
where $P^i_\alpha$ are homogeneous functions of 
degree $3/2$ of $y^1 \cdots y^m$.
Hence the superficial degree of divergence is negative
on this $10m$-dimensional hyper-plane.
Therefore the superficial degrees of divergences are negative
on any hyper-planes in ${\bf R}^{10(N-1)}$.
This completes the proof that the theory is infrared finite
for finite $N$ to all orders. Since the theory is manifestly
finite at short distances, it also establishes that the IIB matrix
model is a finite theory for finite $N$.


Finally we give concrete examples which illustrate the fact that
the factor $(x^i-x^j)^{-3/2}$ can be associated with every $\xi^i-\xi^j$.
At the one-loop level, the diagrams are made of
the bosonic propagators (\ref{eq:bosonprop}),
the fermionic propagators (\ref{eq:fermionprop})
and the $\xi$ insertion vertex (\ref{eq:xivertex}).
The Feynman diagrams contain them consecutively
along the loop as  depicted in the following figure:
\vspace{-1.5cm}
\beq
\begin{picture}(200,80)(60,40)
\GCirc(130,70){4}{1}
\GCirc(105,50){4}{1}
\GCirc(105,20){4}{1}
\GCirc(130,5){4}{1}
\GCirc(165,5){4}{1}
\GCirc(190,20){4}{1}
\GCirc(190,50){4}{1}
\GCirc(165,70){4}{1}
\ArrowLine(129,66)(109,50)
\ArrowLine(105,54)(126,70)
\Photon(108,47)(108,23){2}{3}
\Photon(103,23)(103,47){2}{3}
\ArrowLine(109,20)(129,9)
\ArrowLine(127,4)(106,17)
\Photon(134,7)(163,7){2}{3}
\Photon(163,3)(134,3){2}{3}
\ArrowLine(167,8)(187,21)
\ArrowLine(189,17)(167,4)
\Photon(193,48)(193,24){2}{3}
\Photon(188,24)(188,48){2}{3}
\ArrowLine(169,70)(190,54)
\ArrowLine(187,51)(166,66)
\Photon(133,72)(162,72){2}{3}
\Photon(162,68)(133,68){2}{3}
\Line(128,2)(123,-5)
\Line(168,2)(172,-5)
\Line(102,18)(98,13)\Line(102,52)(97,56)\Line(128,73)(125,78)\Line(167,73)(172, 
77)\Line(194,53)(197,57)\Line(193,18)(198,13)
\Text(148,40)[]{$i$}
\Text(215,40)[]{$j$}
\end{picture}
=tr S^4_{(ij)}.
\vspace{1.5cm}
\eeq
Therefore we can diagrammatically understand that
we can precisely assign the factor $(x^i-x^j)^{-3/2}$
to $(\xi^i-\xi^j)$  in the one loop effective action.

Since the expansion parameter is $g^2$ over
the fourth power of the average distances between 
the points,
the infrared convergence is expected to become better
in higher orders.
In fact at higher loop level, there  remain some vertices
with nontrivial $X$ dependencies which make the infrared convergence
property even better after absorbing half of the powers of the bosonic and the
fermionic propagators connected to every $\xi$.
For example,
\begin{equation}
\begin{picture}(200,50)(-50,20)
\Photon(12,62)(35,43){2}{3}
\Photon(35,43)(59,64){2}{3}\Photon(59,59)(40,40){2}{3}
\Photon(40,40)(60,21){2}{3}\Photon(58,16)(35,36){2}{3}
\Photon(35,36)(13,16){2}{3}
\Photon(12,21)(30,40){2}{3}\Photon(30,40)(12,57){2}{3}
\Text(35,60)[]{$i$}
\Text(54,40)[]{$j$}
\Text(18,40)[]{$k$}
\GCirc(62,60){4}{1}
\GCirc(62,18){4}{1}
\GCirc(8,18){4}{1}
\GCirc(8,60){4}{1}
\ArrowLine(60,21)(60,56)
\ArrowLine(10,56)(10,21)
\ArrowLine(65,56)(65,21)
\ArrowLine(5,21)(5,56)
\Line(64,64)(69,69)
\Line(65,15)(70,10)
\Line(6,15)(2,11)
\Line(6,64)(2,69)
\end{picture}
\end{equation}
is proportional to 
$(\xi^{ij})^{2}/(x^{ij})^{3}\cdot(\xi^{ik})^{2}/(x^{ik})^{3} \cdot g^2
/((x^{ij})^{2}(x^{ik})^{2})$, where the third factor is associated with
the four gluon vertex at the center of this diagram
and indeed makes the infrared convergence property better.
Here we have introduced the notation
$x^{ij}_\mu=x^i_\mu-x^j_\mu$ and 
$\xi^{ij}_\alpha=\xi^i_\alpha-\xi^j_\alpha$.

\subsection{Short distances}\label{sec:short distance}

Until now we have considered the effective action for the diagonal
elements which is valid when all of the diagonal elements are
well separated from one another. It is analogous to the most generic
moduli space in the super Yang-Mills theory although precisely speaking
there is no moduli in the IIB matrix model.
It is natural to consider the second most generic moduli space next.
Let us suppose that a pair of the bosonic coordinates are degenerate
but the rest of the coordinates
are well separated from one another and from the center of mass
coordinates of the pair.

We then decompose the $N \times N$ matrix valued variables
into the $2 \times 2$ submatrices and the rest.
The rest consists of the Hermitian $(N-2) \times (N-2)$ matrices and
the complex $2 \times (N-2)$ matrices.
Let us further decompose the
$2 \times 2$ submatrices into the center of mass part and the
remaining $SU(2)$ degrees of freedom.
Since the off-diagonal elements of $(N-2) \times (N-2)$ matrices
and the complex $2 \times (N-2)$ matrices are massive, we can integrate
them out as before.
We write the remaining degrees of freedom as
\beqa
A_{\mu} &\ni&
\left(\begin{array}{cccc}
A^{U(2)}_{\mu}&&&\\
&x^3_{\mu}&&\\
&&\ddots&\\
&&&x^N_{\mu}
\end{array}\right),\qquad
A^{U(2)}_{\mu}=x_{\mu}{\bf 1} + A^{SU(2)}_{\mu},\\
\psi &\ni&
\left(\begin{array}{cccc}
\psi^{U(2)}&&&\\
&\xi^3&&\\
&&\ddots&\\
&&&\xi^N
\end{array}\right),\qquad
\psi^{U(2)}=\xi{\bf 1} + \psi^{SU(2)}.
\eeqa

The resulting effective action can be written as
\beqa
&&S_{\rm eff}[A^{SU(2)}_{\mu},\psi^{SU(2)};x_{\mu},\xi;
x^3_{\mu},\cdots,x^N_{\mu},\xi^3,\cdots, \xi^N]\n
&=&S[A^{SU(2)}_{\mu},\psi^{SU(2)}]
+S'_{\rm eff}[x_{\mu},\xi;x^3_{\mu},\cdots,x^N_{\mu},\xi^3,\cdots, \xi^N]\n
&&+S_{\rm int}[A^{SU(2)}_{\mu},\psi^{SU(2)};x_{\mu},\xi
;x^3_{\mu},\cdots,x^N_{\mu},\xi^3,\cdots, \xi^N].
\label{eq:effsu2}
\eeqa
The first term is the $SU(2)$ part of the original action.
The second term is essentially identical to the
$SU(N-1)$ case we have discussed in this section.
The center of mass coordinates of the degenerate pair play the
$(N-1)$-th coordinates.
The only difference is that the strength of the
potential between the center of mass coordinates and the others
has doubled.
The third term denotes the interaction between $SU(2)$ part
and the diagonal elements of $N-2 \times N-2$ matrix.
We neglect it since it is small compared to the first term
in many cases as is discussed in the concluding section.

We still need to consider the $SU(2)$ part to determine the
dynamics of the relative coordinates of the pair of the points.
Here we can cite the exact solution for the $SU(2)$ case.
As we show in the appendix \ref{sec:su2},
the distribution for the relative coordinates $r_{\mu}$ is
\beqa
&&\int d^{10}r f(r),\n
&& f(r) \sim \left\{ \begin{array}{ll}
                   1/r^{24} & r^2 \gg g \\
                   r^8      & r^2 \ll g
                \end{array}\right. ,
\eeqa
We conclude that there is a pairwise repulsive potential
of $-8\:\ln r $ type
when two coordinates are close to each other.
It is clear that these considerations are valid
for arbitrary numbers of degenerate pairs although
the center of mass coordinates should be well separated.
Although it is possible to repeat these considerations
to the cases with higher degeneracy,
the analysis becomes more complicated.
Therefore we choose to adopt a phenomenological approach
and assume the existence of the hardcore repulsive potential
of the following form:
\beq
S_{\rm core}[X] = \sum_{i < j} g(x^i-x^j),
\label{score}
\eeq
where
\beq
g(x^i-x^j)=\left\{ \begin{array}{ll}
                    -4 \ln ((x^i-x^j)^2/g) & {\rm for}\;(x^i-x^j)^2 \ll g \\
                    0                      & {\rm for}\;(x^i-x^j)^2 \gg g
                    \end{array}\right. .
\eeq
In the following section, we investigate the structure of space-time
by using the one loop effective action for the diagonal elements
plus the phenomenological hardcore potential $S_{\rm core}$.
Although our effective action is not exact at short distances,
it presumably captures the essential feature of the IIB matrix model
namely the incompressibility.
If so our effective action may be in the same universality class of the
full IIB matrix model.

\setcounter{equation}{0}
\section{Possible scenarios of dynamical generation of space-time}
\label{sec:section3}
So far we have derived an effective action for
the diagonal elements of  ${A_{\mu}}$ and $\psi$.
We have identified the eigenvalues of $A_{\mu}$
with the space-time coordinates $X_{\mu}$ through the semiclassical
correspondence as is explained in the introduction.
Let us suppose that the eigenvalue distribution of $A_{\mu}$ is
extensive in $d$ dimensions but supported only
by finite intervals in the remaining $10-d$ dimensions.
Such a distribution may be interpreted that space-time is
extensive in $d$ dimensions but has shrunk in the
remaining directions. The dimensionality of such
space-time is obviously $d$.
If such a distribution is derived from our effective action,
we may conclude that the dimensionality of space-time is $d$.

In the perturbative formulation of string theory,
we need to assume a particular background on which string propagates.
Although certain consistency conditions are required for the
backgrounds, we have no principle to pick a particular background
from the others. However we can in principle determine
the possible structure of space-time from the IIB matrix model uniquely.
Our effective action provides us an exciting possibility to
realize it.

In this section  we study the effective action
in order to see whether the IIB matrix model can explain
the structure of space-time and its dimensionality in particular.
As is explained shortly in more detail, our effective action is still
rather complicated. 
One possible way to avoid it is to consider
further simplified models which are supposed to contain
the essential features of our effective action.
Such models may contain several parameters
which may not be easily determined.
We therefore draw the phase diagrams of those models
for all possible parameter regions and  identify the structures of
space-time in the respective phases.

The aim here is to show the existence of models
which can realize four dimensional space-time.
We then need to show that the IIB matrix model
indeed belongs to the same universality class.
Although these investigations are still in progress,
we find it very plausible
that the IIB matrix model realizes four dimensional flat space-time.
We explicitly propose some of possible mechanisms to realize realistic
space-time.


\subsection{Models}\label{sec:model}
We consider the following ensemble in ten dimensions
which is controlled by the 1-loop effective action
$S_{{\rm eff}}^{{\rm 1-loop}}[X, \xi]$ (\ref{eq:seff}) and
the core potential (\ref{score}):
\beqa
&& \int dX e^{-S_{\rm core}} d\xi e^{-S^{\rm 1-loop}_{\rm eff}[X,\xi]}
\nonumber \\
& = &\int dX e^{-S_{\rm core}} d\xi \sum_{G:{\rm graph}}\; \prod_{(ij):{\rm bond\;of\;}G}
               [(\frac{tr(S_{(ij)}^4)}{4})\;\; {\rm or}\;\;
                (\frac{1}{2}(\frac{tr(S_{(ij)}^4)}{4})^2
                +\frac{tr(S_{(ij)}^8)}{8})].
\label{ST1}
\eeqa
As is explained in section \ref{sec:effdia},
we call the bond in association with
$ tr (S_{(ij)})^4 $ and the bond with
$ (tr (S_{(ij)})^4)^2 /32+   tr (S_{(ij)})^8 /8 $
a 8-fold bond and a 16-fold bond respectively.
The former gives 8 fermionic variables per bond while
the latter gives a product of all 16 components
and is proportional to $ (x^{ij})^{-24} \delta^{(16)}(\xi^i -\xi^j) $.
\par
If only 16-fold bonds appear in a graph,
$\xi$ integration can be easily performed which gives
a branched polymer as we saw in the previous
section. The integrations can be performed bond by bond
and the resultant interactions between $x^i$'s are functions of
distances only, that is,
 $(x^i-x^j)^{-24}$ for each bond.
Thus we can estimate the partition function as
\beq
\int dX e^{-S_{\rm core}} d\xi e^{-S^{\rm 1-loop}_{\rm eff}[X,\xi]}
\longrightarrow
\int dX e^{-S_{\rm core}}
\sum_{G:{\rm maximal \; tree} }\;  \prod_{(ij):{\rm bond\;of\;}G}
 {1 \over (x^i-x^j)^{24}} ,
\label{ST2}
\eeq
where summation of $G$ is over all maximal tree graphs.
\par
When 8-fold bonds appear in the graph, 
the  simple branched polymer picture breaks down.
As is seen from eqs. (\ref{eq:sbt}) and (\ref{tmn}),
$tr (S_{(ij)})^4$ can be rewritten as
\begin{equation}
tr (S_{(ij)}) ^4 \propto
 C^{\mu \nu \lambda \rho \alpha_1 ...\alpha_8}
\xi^{ij}_{\alpha_1} \cdots \xi^{ij}_{\alpha_8}
V^{ij}_{\mu \nu \lambda \rho}/(x^{ij})^{16},
\label{ST3}
\end{equation}
where $\xi^{ij}_\alpha=\xi^i_\alpha - \xi^j_\alpha$,
$x^{ij}_{\mu}=x^i_\mu-x^j_\mu$,
and $C^{\mu \nu \lambda \rho \alpha_1 ...\alpha_8}$ is an invariant
tensor. $V^{ij}_{\mu \nu \lambda \rho}$ is a fourth rank
symmetric traceless
tensor constructed from $x^{ij}$:
\begin{eqnarray}
V^{ij}_{\mu \nu \lambda \rho}
&=& x^{ij}_{\mu} x^{ij}_{\nu} x^{ij}_{\lambda} x^{ij}_{\rho} -
{(x^{ij})^2 \over D+4}(x^{ij}_{\mu} x^{ij}_{\nu} \delta_{\lambda \rho}
+ x^{ij}_{\mu} x^{ij}_{\lambda} \delta_{\nu \rho} + \cdot  \cdot \cdot)
\nonumber \\
&& + {(x^{ij})^4 \over (D+2)(D+4)} (\delta_{\mu \nu} \delta_{\lambda \rho}
+  \delta_{\mu \lambda } \delta_{ \nu \rho} + \cdot  \cdot \cdot) ,
\label{ST5}
\end{eqnarray}
where $D=10$.
Therefore at each 8-fold bond there are
many choices to select 8 fermionic components out of 16 and
each choice gives different $x^{ij}_{\mu}$ dependence
which is specified by $C^{\mu \nu \lambda \rho\alpha_1 ...\alpha_8}$.
Hence integration over $\xi$ generally
gives complicated interactions which involve many bonds and
depend on relative directions of relevant bonds.
An important point here is that the tensor
$V^{ij}_{\mu \nu \lambda \rho}$ is traceless
and vanishes if we naively take the orientation average for $x^{ij}_\mu$.
Therefore it is natural to expect that the appearance of 8-fold bond is 
suppressed to some extent
if the eigenvalues of $A_{\mu}$ are distributed uniformly in ten dimensions.
\par

As a simpler model which may capture the essential feature
of our effective action, we may consider a model where
we average the orientation dependence of the 8-fold bonds.
Instead of considering the graphs consisting of 8-fold and 16-fold bonds,
we introduce two independent maximal trees each of which maximally
connects the $N$ points with coordinate $x^i_\mu$.
When the two trees share a bond $(ij)$, we regard it as a 16-fold bond
and give the corresponding factor $((x^{ij})^{2}+g)^{-12}$.
For a bond which is contained in one tree only, 
we regard it as a 8-fold bond and 
assign a factor
$((x^{ij})^2+g)^{-6}$ which represents the residual 
effect of the 8-fold bond after  
orientation average.
Here we have introduced a short distance
cut-off of order $g$ in order to avoid the meaningless singularity at $x^{ij}=0$.
We also assume the hardcore repulsive potentials as given by $S_{\rm core}$
at short distances:
\beqa
\int dX e^{-S_{\rm core}} d\xi e^{-S^{\rm 1-loop}_{\rm eff}[X,\xi]}
&\longrightarrow&
\int dX e^{-S_{\rm core}}
\sum_{G:{\rm two\; maximal\; trees}}\;    \prod_{(ij):{\rm bond\;of\;}G}\n
&&\times[ ((x^{ij})^2+g)^{-6}e^{-\lambda /2}
  \;\; {\rm or}\;\;  ((x^{ij})^{2}+g)^{-12}   ] .
\label{ST7}
\eeqa
Here the meaning of the factor $e^{-\lambda/2}$ is as follows.
As we have discussed in the previous paragraph, the orientation average
suppresses the appearance of 8-fold bonds.
Therefore we  can
take the effect of angle-dependence at least partially by introducing a 
suppression factor $e^{-\lambda/2}$ for each 8-fold bond.
We call this model a ``double tree model''.

If the IIB matrix model is really approximated by this model,
$\lambda$ should be in principle fixed 
because the IIB matrix model has no free parameter.
However to determine $\lambda$ precisely may be  as difficult as 
solving the IIB matrix model itself.
Therefore we treat $\lambda$ as a parameter,
and examine the phase structure as $\lambda$ is varied form 
$-\infty$ to $\infty$.
We have employed numerical simulations.
We show in the next subsection \ref{transition}
that two apparent phases exist in two opposite limiting cases
$\lambda \rightarrow \infty$ and $\lambda \rightarrow -\infty$ .
\par
An important factor which is not  taken into account
in the double tree model
is the fact that $\lambda$ is a sort of mean field and it  
may strongly depend on the state of the system, that is,
the distribution of eigenvalues.
Actually
we will see in subsection \ref{angle dep} that $\lambda$
decreases
when the eigenvalues distribute in lower dimensional space-time,
and this effect favors to generate lower dimensional space-time.
By introducing a rough approximation we can show that
the free energy is minimized if the space-time is compactified to 
four dimensions.
It appears that more detailed analysis of this effect is indispensable
to arriving at our goal.

\subsection{Branched polymer phase and droplet phase \label{transition}}

In this subsection we analyze the double tree model
defined by  eq.(\ref{ST7}).
It contains a parameter $\lambda$ and
we consider two limiting situations.
First we take $\lambda$ large enough. 
Then most of the bonds on one tree are bound to bonds on the other tree
and
the system  behaves as an ordinary
branched polymer made of 16-fold bonds only.
We call this state a branched polymer (BP) state.
As is reviewed in appendix \ref{app:BP}, its partition
function is given by
\beqa
&& Z _{\rm BP}(N) = N! (\hat{f_1}(0) \alpha_c)^N 
 \sim N! \
e^{-N F_{\rm BP}} ,
\nonumber
\\
&& \hat{f_1}(0) = \int_{x_c} d^D x {1 \over x^{24}} ,
\label{ST10}
\eeqa
where we have introduced free energy per bond defined by
$F_{\rm BP}= -\ln (\alpha_c \hat{f_1}(0)) $.
If we neglect the effect of intersections
due to the core potential, we can set $\alpha_c =e$.
We have also multiplied it by $N!$ since we
distinguish $N$ points $x^i$.
$x_c$ is a short distance cut-off determined by the
core potential.
Since the Hausdorff dimension of BP is four,
we cannot neglect the effect of intersections
in dimensions less than eight and
eq. (\ref{ST10}) gives an overestimation of the partition function.
In other words, $\alpha_c$ is reduced from $e$.
Furthermore in dimensions
less than four most of  BP's  cannot be packed and the number of
possible graphs drastically decreases.
Without any other interactions besides 16-fold bond interactions,
BP favorably expands in ten dimensions and we cannot get
lower dimensional distribution of eigenvalues.
\par
On the other hand if we take $\lambda$ small, the number of 8-fold bonds
increases and 16-fold bonds totally disappear in the 
$\lambda \rightarrow -\infty$
limit. 
Actually even in some positive values of $\lambda$
most of 16-fold bonds disappear,
since the entropy  increase caused by resolving
16-fold bonds into 8-fold bonds overcompensates the effect of 
positive values of $\lambda$
which favors 16-fold bonds.
Our numerical simulation supports this fact as we will see later.
Then the  entropy is gained for such a configurations that
as many as possible
points gather around one another
so  that rearrangements of bonds occur easily.
Indeed, without core potential
all points are condensed into a finite area whose size does not depend on $N$,
as we show at the end of this subsection.
If we take the core potential into account,
they cannot condense into an infinitely dense state.
Instead
all of the cores are packed in an area whose size grows with $N$.
We call this state a droplet state.
When a droplet state
is realized in $d$ dimensions, the size of the droplet
is given by $R= l (N)^{1/d}$, where $l$ is the  core size.
Since bonds flip locally in a droplet, we regard the droplet
as a collection of 
clusters in each of which bonds can freely flip.
The cluster is assumed to have $n$ points inside.
Then the volume of the cluster is $v=l^d n$,
and the typical length of bonds in the cluster can be identified with
the radius of the cluster $r=l n^{1/d}$.
Therefore each cluster is expected to contribute to the 
partition function by 
\beq
z(n) = \left( n! \left( {\alpha_c \over r^{12}} \right)^n  \right)^2 v^n
e^{-\lambda n_8/2},
\eeq
where $n_8$ is the number of 8-fold bonds in the cluster.
The first factor is the partition function of BP consisting of $n$ points, and
is squared since we sum over two
independent maximal trees.
We can then roughly estimate the partition function of
the double tree model in the droplet phase by
\begin{eqnarray}
   Z_{\rm droplet} &\sim&
 (z(n))^{N/n} {N! \over (n!)^{N/n}}
  \nonumber \\
 & \sim & N! n^{2N (1-12/d)} (\alpha_c^{2} \ l^{d-24})^N e^{-\lambda N_8/2},
\label{ST20}
\end{eqnarray}
where $N_8$ is the total number of the 8-fold bonds.
Since $d \le 10$, $Z_{\rm droplet}$ is maximized by a small value of $n$ which
should be independent of $N$.
Therefore $Z_{\rm droplet}$ can be written as
\begin{eqnarray}
Z_{\rm droplet} \sim  N! \ e^{-N F_{\rm droplet}-\lambda N_8/2}.
\label{ST30}
\end{eqnarray}

\begin{figure}
\epsfxsize = 8 cm
\epsfysize =  6 cm
\centerline{\epsfbox{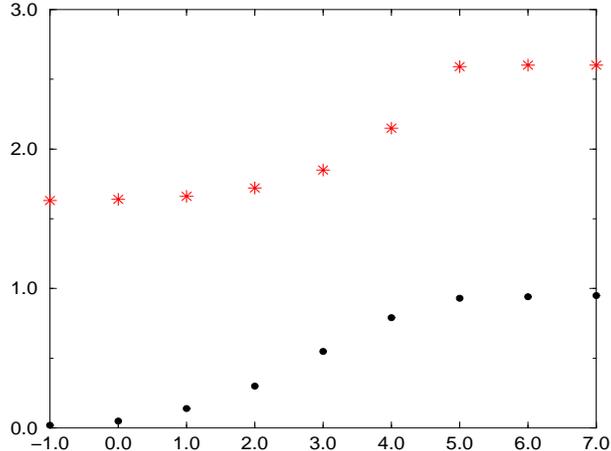}}
\caption{Numerical simulation for $N=1000$ .
Horizontal axis is $\lambda$. Stars indicate
root mean squares of the eigenvalues and dots indicate the ratio of
16-fold bonds. For large values of $\lambda$, the system is in a BP phase.
For small values of $\lambda$, it is in a droplet phase. }
\label{fig:transition}
\end{figure}

Comparing the partition function of BP eq.(\ref{ST10})
and that of droplet eq.(\ref{ST30}), we can tell that
there is a phase transition or a cross-over between the two phases at
some positive  value of $\lambda$.
For larger values of $\lambda$, the system is in a BP phase and
for smaller values of $\lambda$, it is in a droplet phase.
Our numerical simulation supports this transition
(see fig. \ref{fig:transition}).
The nature of the transition is not clear at the present stage,
although it has a similarity to the liquid-gas transition
since we have no order parameter associated with a symmetry.
The points distribute in full ten dimensions both in these two phases.
Other possible interactions such as
angle dependent interactions  will  modify these
phases and we may obtain a phase with lower dimensionality.
\par
Finally in this subsection we see what happens if we do not have the
core potential. 
In this case
we can use mean field approximation to determine one-particle
density in the following way.
Let us denote
$p_n$ as the probability that a point is connected to $n$ other points.
Then the self consistent equation for one-particle density is given by
\begin{eqnarray}
&& \rho(x) = {1 \over Z} \exp \left(\sum_{n=1}^{\infty} p_n
n \int \ln (f(x-y)) \rho(y) d^d y \right),
\nonumber \\
&& \int \rho(x) d^d x =1,
\label{ST50}
\end{eqnarray}
where $f(x)$ is the Boltzmann factor for each bond and given by
$(x^2+g)^{-6}$ in our case. 
Since $N$ does not appear in these equations, the
one-particle density does not depend on  $N$.
Hence all points are condensed into a finite area $V$ which does not 
depend on $N$.
We call this state a mean-field state.
We can find 
a lower bound of the partition function by
replacing the length of each bond with the size of the whole system:
\begin{equation}
Z_{\rm MF} > ( N! \alpha_c^N)^2 (V^{-12N/d})^2 V^{N} e^{-\lambda N_8 /2}.
\label{ST60}
\end{equation}
Since eq. (\ref{ST60}) contains extra $N!$ compared with $Z_{\rm BP}$ in
eq. (\ref{ST10}), we have to increase $\lambda$ as
$\log N \rightarrow \infty$ in order to keep the state in BP phase.
Thus we find that any state falls into a mean-field phase
if we do not have the core potential.
\subsection{A mechanism favoring lower dimensions \label{angle dep}}
Although the double tree model has taught us a lot about the IIB matrix model,
we may need to extend it in order to be more faithful to our effective action.
As we saw in subsection \ref{sec:model}, the interaction of 8-fold
bonds generally depends on the relative angles of the vectors
$x^{ij}_{\mu}$.
If eigenvalues  extend in  ten dimensions, averaging over directions 
gives a suppression factor to each 8-fold bond,
because the tensor $V_{\mu \nu \lambda \rho} $ is traceless.
This suppression becomes weak
if eigenvalues collapse into lower dimensions.
Therefore if there are considerable number of 8-fold bonds,
the partition function is suppressed in higher dimensions. 
However the entropy which is associated
with possible orientations of bonds favors higher
dimensions.
Therefore there is a competition between the two
and it is conceivable that four dimensional space-time
is realized as a compromise.
In what follows we roughly estimate the partition function 
assuming that the system is in a droplet phase
and show the possibility of this scenario.


Assuming that $x^{ij}_{\mu}$ lie isotropically in a $d$ dimensional 
subspace, we can estimate the orientation average of 
$V^{ij}_{\mu \nu \lambda \rho}$  as
\begin{eqnarray}
V^{ij}_{\mu \nu \lambda \rho}
&\sim&
\left({2 (x^{ij})^4 \over d (d+4)}-{(x^{ij})^4 \over (d+2)(d+4)}\right)
(\hat{\delta}_{\mu \nu} \hat{\delta}_{\lambda \rho}
+\hat{\delta}_{\mu \lambda}
\hat{\delta}_{\nu \rho} + \cdot  \cdot \cdot)\nonumber \\
&&
 -{ (x^{ij})^4 \over d (D+4)}(\hat{\delta}_{\mu \nu} {\delta}_{\lambda \rho}
+\hat{\delta}_{\mu \lambda} {\delta}_{\nu \rho} + \cdot  \cdot
\cdot)\nonumber \\
&& + {(x^{ij})^4 \over (D+2)(D+4)} (\delta_{\mu \nu} \delta_{\lambda \rho}
+  \delta_{\mu \lambda } \delta_{ \nu \rho} + \cdot  \cdot \cdot) .
\label{ST105}
\end{eqnarray}
Here $\hat{\delta}$ stands for the projection operator to the 
$d$ dimensional space-time  while $\delta$ is 
the Kronecker's delta in the original $D=10$ dimensions.
The interactions of 8-fold bonds in general involve
many  $V_{\mu\nu\lambda\rho}$'s
whose Lorentz indices are contracted in
various ways.
It is not easy to estimate the combinatorics which come out
from such contractions exactly even after such simplification
as eq. (\ref{ST105}). It may be qualitatively valid to
replace $\hat{\delta}$ by ${\delta}$. Then the relevant combinatorics
become independent of $d$.
Under such an approximation, each
8-fold bond becomes to have a wight effectively which is proportional to 
\begin{equation}
e^{-\lambda/2} \sim 
f_{\rm eff}(d)= \left( {1 \over (D+2)(D+4)} - {1 \over (d+2)(d+4)}
+ {2 \over d} \left( {1 \over d+4} -{1 \over D+4} \right) \right) .
\label{ST110}
\end{equation}

\begin{figure}
\epsfxsize = 8 cm
\epsfysize =  6 cm
\centerline{\epsfbox{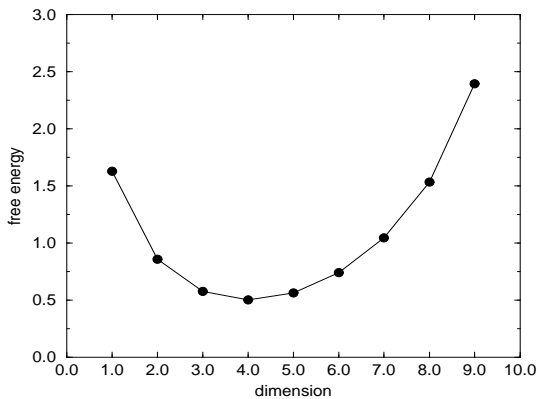}}
\caption{Free energy as a function of dimensions.
We take $\gamma=17/6$. Free energy is minimized at four dimensions.
}
\label{fig:angle}
\end{figure}

The weight $f_{\rm eff}(d)$ is a decreasing function
and favors lower dimensions. Competition between this weight
and entropy which favors higher dimensions determines
space-time dimensions in this scenario.
If we assume that all bonds are 8-fold and 
the entropy per bond increases as $d^{\gamma}$, the partition
function for a droplet can be expressed as
\begin{equation}
Z_{\rm droplet} = N! e^{-2 N F(d)} \sim N! (f_{\rm eff}(d))^{2N} d^{2\gamma N}.
\label{ST120}
\end{equation}
The free energy $F$ has a minimum at $d=4$ when $\gamma$ is
between 2.6 and 3.1.
Fig. \ref{fig:angle} shows $F$ as a function of $d$.
If this is the mechanism to generate our space-time, we need to
determine detailed numerical factors to prove it.
We stress here that the model is already fixed by nature
and we don't have freedom to tune parameters.

\subsection{Other mechanisms for understanding space-time structure}
There are two other possible mechanisms to generate lower
dimensional  structure of
the space-time.
\par
One way is to approach from a BP phase in which 
we regard 8-fold bonds in \eq{ST1} as 
perturbations to BP's consisting of 16-fold bonds only.
Since Hausdorff dimension
of BP is four, there is a possibility
that small perturbations might compress the system into four
dimensions. If we neglect the effect of self-intersections and naively confine
 BP into $d$ dimensional space,
mean density becomes zero for $d>4$ and diverges for $d<4$
in large $N$ limit.
Then for $d<4$, the core potential extends the bond length
and energy drastically increases.  In other words most of the topologies of BP
are prohibited to avoid the increase of energy and the  entropy
is decreased very much.
For $d>4$, large distances between the points prevent 16-fold bonds from 
resolving into pairs of 8-fold bonds, and
we cannot gain the entropy of having 8-fold bonds at various places on 
the polymer.
On the other hand at $d=4$, 8-fold bonds can move around on the polymer
and the system behaves as a gas of 8-fold bonds.
Therefore only 
at $d=4$ we can gain the entropy of 8-fold bond gas without too much
excess of energy due to the core potential.
The mechanism discussed here is based on the picture of BP 
and seems totally different from the one discussed in subsection \ref{angle dep}
which is based on the picture of droplet.
In four dimensions, however, these two pictures can be complementary,
since the size of space-time predicted in these pictures are 
both $N^{1/4}l$ and agree with each other.
\par
Finally we point out a possibility that four-dimensional space-time is 
realized in the intermediate region of fig. \ref{fig:transition}
even for the naive double tree model where dimension dependence of 
$\lambda$ is not taken into account. 
Here we consider dimension
dependence of the ratio of 8-fold bonds, $r_{8}$.
In subsection \ref{transition} we saw the transition between
two limiting phases, namely
a BP phase for large
$\lambda$ and a droplet phase for small $\lambda$.
For intermediate values of  $\lambda$,
the partition function can be expressed as 
\begin{equation}
Z(\lambda) \sim N! \ e^{-N(F(\lambda) + \lambda r_{8})},
\label{ST130}
\end{equation}
where $F(\lambda=\infty)= F_{\rm BP}$ and $F(\lambda= -\infty)=F_{\rm droplet}$.
Obviously the free energy $F(\lambda)$ defined above
is always smaller in higher dimensions.
Therefore the system always expands in full dimensions,
if $r_{8}$ is independent of dimensions.
Our numerical simulations, however, showed  that
$r_{8}$  is smaller in lower dimensions
for fixed values of $\lambda$.
It is consistent with the intuition that 
each point has a smaller number of neighbors in lower dimensions
and the probability that two 8-fold bonds coincide 
to form 16-fold bond is larger.
If the dimension dependence of $\lambda r_{8}$ exceeds that of $F(\lambda)$,
 there is a possibility that lower dimensional phase  becomes
stabler than the ten dimensional phase for some value of $\lambda$.
%
%

\section{Conclusions and Discussions}
\setcounter{equation}{0}

In this concluding section, we first estimate the magnitude
of the higher loop contributions to
the effective action.
Since our expansion parameter is $g^2$ over the fourth power
of the average distances between the points,
long distance behavior is well described by
the one-loop effective action $S_{\rm eff}^{\rm 1-loop}[X]$.
However there is a possibility that small contributions may
pile up at each loop since $N$ color indices circulate around
the loop.
Our strategy in this section is to assume that the one loop
level effective action determines the distribution $\rho (x)$
of the eigenvalues. We then use $\rho (x)$ to estimate the magnitude
of the higher loop effects. This procedure is self-consistent if
we find that the higher loop effects are small.
Of course, our perturbative expansion breaks down if the
expansion parameter grows as a positive power of $N$. 
What we show
here is that the higher loop corrections are indeed finite in the large $N$ limit
for finite coupling constant,
and our approximation based on the loop expansion can be fully justified.

Although we have argued that higher loop effects
improve the infrared convergence properties in general,
there are diagrams which possess the identical
infrared convergence properties with the one loop
diagrams in our estimation procedures just explained.
For example  we may consider the one loop
self energy corrections to the propagators of the off-diagonal
components.
We then consider the insertions of them into the
propagators of the
one loop diagrams which contribute to the effective action.
It is a  part of the two loop level contributions.

Recall that the matrix model (\ref{action}) has been
constructed by the large $N$ reduction procedure from the $SU(N)$ gauge theory.
Thus, in the leading order in $1/N$ expansion,
the perturbative expansions in $g^2$ in the matrix model are identical to
those in the gauge theory were it not for the fermionic zeromodes.
They are precisely the part of the quantum corrections we
quoted above.
Then the ratio of $(l+1)$-loop contributions to $l$-loop contributions may
be estimated by dimensional analysis as
\beqa
R_{g^2}
&=&g^2\sum_k\frac{1}{(x^i-x^k)^4} \n
&=&g^2 N\frac{\sum_k (x^i-x^k)^{-4}}{\sum_k 1}\n
&=&g^2 N \frac{\int d^{10}x\; \rho (x) x^{-4}}{\int d^{10}x\; \rho (x)},
\label{eq:rg2def}
\eeqa
where $\rho (x) =\sum_k \delta(x-x^k)$
is a single-particle distribution of the eigenvalues.

We first assume that $x$ is distributed uniformly
in a $d$ dimensional manifold with size $R$.
From the correspondence with gauge theories, we can say that
there is no divergences from small $x$ regions in $d>4$,
and from large $x$ regions in $d<4$.
Thus,
\beq
\int^R d^dx\;x^{-4}=
\left\{\begin{array}{ll}
R^{d-4} & d>4 \\
l^{d-4} & d<4
\end{array}\right.,
\eeq
where $l$ is a typical size of short distance cutoff.
Therefore,
\beq
R_{g^2} =
\left\{\begin{array}{ll}
\frac{g^2 N}{R^4} = N (\frac{l}{R})^4 & d>4\\
\frac{g^2 N}{R^d l^{4-d}} = N(\frac{l}{R})^d & d<4
\end{array}\right.,
\eeq
where we used the fact that the core size is $l \sim g^{1/2}$.
Since $N \sim (R/l)^d$, we may conclude that the quantum corrections are
divergent for $d >4$ but finite for $d<4$ in the large $N$ limit.
These arguments seems to imply the logarithmic divergence for $d=4$.
However here we need to take into account of supersymmetry.
It is well-known that dimensional reduction from ten dimensional
super Yang-Mills theory down to four implies ${\cal{N}}=4$ supersymmetry
in four dimensions. Furthermore four dimensional super Yang-Mills theory
with ${\cal{N}}=4$ supersymmetry is also known to be finite.
So we expect that the quantum corrections for $d=4$ case is finite.

We have found a divergence in the above argument for $d>4$.
However the situation is totally different if we assume that
space-time is branched polymer like for $d>4$.
We recall that $R\sim N^{1/4} l$ for such space-time
since its Hausdorff dimension is four.
If so, the estimation of \eq{eq:rg2def} is essentially the same as
the four dimensional case, and
we may conclude that quantum corrections are finite also
for $d>4$.
Thus we can show that the corrections coming from planner diagrams
are finite in the large $N$ limit if the coupling $g^2$ is kept fixed.

Next let us consider higher order effects of $1/N$ expansion.
It is clear that the leading $1/N^2$ corrections
contain two less summations over color indices
compared with the planar contributions.
However they do contribute since they  are at least finite.
In fact they can be of the same orders of magnitude with the planar
contributions since they are more singular at small $x$
regions.
For example  
let us consider a three loop correction
to the gluon self-energy.
Such a nonplanar contribution should go like
\beq
g^6 \sum_k
{1\over (x^i-x^j)^6}{1\over (x^i-x^k)^2}{1\over (x^j-x^k)^4} .
\eeq
Here the important point is that the superficial degree of divergence 
for the integration variable $x^k$ is reduced
compared with that of the corresponding planar diagram.
For dimensions up to four, the above expression may be
evaluated as follows:
\beqa
&&g^6 N {\int  d^{10} x\rho (x)
{1\over (x^i-x^j)^6}{1\over (x^i-x)^2}{1\over (x^j-x)^4}
\over \int  d^{10} x\rho (x) }
\nonumber \\
&=& g^6 N {1\over l^{12-d} R^d} ,
\eeqa
where we set $(x^i-x^j)=l \sim g^{1/2}$ to estimate the 
magnitude of the correction at short distances.
Since  $N \sim (R/ l)^d$,
the contribution  is again nonvanishing and finite in the large $N$ limit.

Therefore infinite numbers of diagrams  contribute
in dimensions up to four.
This means that all orders of $1/N$ expansion contribute equally.
On the other hand the nonplanar contributions
diverge in general for $d>4$ although they are
much smaller compared with the planar contributions
unless space-time is  branched polymer like.
If the space-time is branched polymer like, the situation is essentially
identical to the four dimensional case since its fractal dimension
is also four. So we may conclude that all orders of $1/N$ expansion
contribute in all dimensions.

Our findings may be interpreted that
$g_{\rm st} \sim 1$ since we have argued that string coupling
constant is inversely proportional to $N$ \cite{FKKT}.
Apparently the double scaling limit is naturally taken in
the IIB matrix model.
However our investigations here also imply that the string scale
$(\alpha ')^2 \sim g^2$. It is in disagreement with
our previous estimate $(\alpha ')^2 \sim g^2 N$ based on the loop equations.
We believe that reexaminations of the loop equations should be
able to reconcile this discrepancy.

We also would like to comment on the universality of IIB matrix models.
It is clear from the investigations in this section that
the renormalizability and finiteness of four dimensional
super Yang-Mills theory is crucial to control the quantum
corrections in the IIB matrix model.
If we were to add higher dimensional terms
to the IIB matrix model, we may need to renormalize the
theory nonperturbatively.
We expect that such a renormalized theory belongs to the same
universality class with ours.
Therefore we may argue that the IIB matrix model is universal
and our Lagrangian may correspond to the fixed point Lagrangian.

In this paper, we have derived an effective action for the IIB matrix model
which is valid when the eigenvalues of $A_{\mu}$ are widely separated.
It is the effective action in terms of the super coordinates of space-time.
It has clearly shown that the theory has no infrared divergences and the
universe never disintegrates.
The exciting possibility is that we can determine the dimensionality of
space-time by studying such an effective action.

There are order $N$ pairwise attractive potentials between the
space-time
coordinates. Such interactions may be classified into the 16-fold bond and
the 8-fold bond types. If there were only the 16-fold bond interactions,
the space-time coordinates form branched polymer type configurations with
the fractal dimension four which stretch in ten dimensional space-time.
Such a distribution of the eigenvalues cannot be considered as
continuous space-time.
We have found that the closely packed distributions of the eigenvalues can be
realized when we take the 8-fold bond interactions into account.
Such distributions can be interpreted as continuous space-time.
Here the existence of the hard core repulsive potentials between the
eigenvalues
is also very important.
Although more detailed investigations are required to show that four
dimensional space-time is realized in this model, we have proposed 
possible mechanisms
to realize the realistic dimensionality.

We believe that what we have already learned from the IIB matrix model
is very significant.
It is not defined on the fixed background metric but space-time is dynamically
determined as the vacuum of this model. The extent and the dimensionality
of space-time can be obtained from the eigenvalue distributions of
$A_{\mu}$.
In this model, the vacuum is not empty and may be classically represented
by the diagonal matrices. The matter (for example D-objects) are the
local fluctuations which float on the vacuum. In this way the space-time
and the matter are inseparable and they determine each other.
Note that such a picture never emerges as long as we deal with the effective
Lagrangians for a finite number of D-objects.

If four dimensional space-time is generated dynamically,
it will be plausible  that the space-time has
four dimensional Lorentz invariance. Also it will be
intrinsically flat.
Here the core potential plays an important role to
protect the space-time from shrinking.
On the other hand long-distance dynamics protects it
from expanding infinitely.
This means vanishing of the cosmological constant.
In our scenario, stability of generated space-time
guarantees absence of cosmological constant in an effective
theory of gravity, that is, Einstein gravity.
If vacuum energy is generated, eigenvalues are rearranged so as to
restore stability of space-time.
It is dynamically tuned to be zero.
The existence of the hardcore repulsive potential at short distances
and the infrared finiteness are responsible to achieve this remarkable feat.
It does not explicitly depend on the supersymmetry and
hence should be independent of whether supersymmetry is spontaneously
broken or not.
We believe that this fact alone constitutes a major
achievement and hence underscores the validity of the IIB matrix model.

We recall the ${\cal N}=2$ SUSY transformations for the zeromodes:
\beqa
\delta^{(1)}\xi &=& 0 ,\n
\delta^{(1)} x_{\mu} &=& i\bar{\epsilon _1}\Gamma_{\mu}\xi ,
\label{Ssymp1}
\eeqa
and
\beqa
\delta^{(2)}\xi &=& \epsilon _2 ,\n
\delta^{(2)} x_{\mu} &=& 0.
\label{Ssymp2}
\eeqa
The linear combinations of the above translations $\tilde{\delta}^{\pm}=
\delta ^{1}\pm \delta ^{2}$
form the ${\cal N}=2$ supersymmetry algebra.
The vacuum expectation value of the supersymmetry transformation
of $\xi$ must vanish if supersymmetry is not broken by the vacuum.
It is clear from eq.(\ref{Ssymp2}) that the ${\cal N}=2$ supersymmetry
of $\tilde{\delta}^{\pm}$ are broken completely.
It is in contrast to the D-string case where the half of the
supersymmetry can be preserved.
If so, this model realizes the vanishing cosmological constant without
supersymmetry as we have argued.

It is also possible to write a scenario to obtain realistic gauge groups
in this model.
Although we have considered the most generic case where all eigenvalues
of $A_{\mu}$ are different from each other,
we may assume that some of the eigenvalues remain degenerate.
Then the
vacuum configurations are represented by the block diagonal matrices where
each block is
$m$ dimensional.
Then the gauge group $SU(m)$ is realized (note that $U(1)$ part is used
to make the space-time coordinates). It is also possible
to realize the standard model gauge group in an analogous way.
We may again integrate out the off-diagonal elements and obtain the
effective action for the $U(m)$ submatrices. The resultant effective
action must closely resemble the gauge theory with the Planck scale
cutoff since we have the
local gauge invariance as the manifest symmetry of the effective action.

Since we have argued that it is possible to obtain the realistic gauge groups
in four dimensions in this model, it is natural to ask to what kind of
string compactification
it corresponds.
It presumably corresponds to the compactification into the interior of $S^5$
that is $B^6$.
It may be possible to play the analogous games with Calabi-Yau
compactifications of heterotic string.
It is possible to obtain chiral spinors in the fundamental
representations of the gauge group in four dimensions from the
chiral spinors in the adjoint representations in ten dimensions.
For this purpose we need to assume that there is nontrivial
gauge field configuration in $B^6$ which produces a nontrivial
index for the Dirac operator.
We immediately think of a two dimensional example
such as magnetic vortices. Then the tensor products of three of them
are the possible candidates for such nontrivial gauge configurations.

We also note the similarity between our effective Lagrangian and the dynamical
triangulation approach for quantum gravity. Due to the existence of the
hardcore potential of the Planck scale, the unit cell of the effective
action is also the Planck scale.  The pairing interaction between the
eigenvalues
may be identified with the bonds in the dynamical triangulation approach.
The integration over the fermionic zeromodes ensures that the model sums the
contributions with all possible connectivities of the bonds just like the
dynamical
triangulation approach.
This analogy may be useful to reconstruct the metric out of the IIB matrix
model.
We also conjecture that the general coordinate invariance is present
in this model due to the following reasoning.
There is the permutation symmetry $S_N$ which permutes the
color indices. It is a subgroup of the full gauge group $SU(N)$
and an exact symmetry of our effective action.
Since it does not change the density of the eigenvalues,
it should be part of the volume preserving diffeomorphism
group in the continuum limit.
Similar argument has been made in the dynamical triangulation approach.
However it suffers from
the infamous conformal mode instability while the IIB matrix model is free
from such a disease which is again a remarkable merit of this model.
So the gauge and the diffeomorphism invariances may be unified into the
$SU(N)$ symmetry of the IIB matrix model.

Although the investigation of this model is still in the initial stage,
this model turns out to be a finite theory which is free from both
short distance and long distance divergences.
The existence of such a theory itself is very impressive.
It is a manifestly covariant formulation of superstring which enables us
to determine the structure of the vacuum, namely space-time.
We have argued that it can solve the problems such as the cosmological constant
problem which appear to be insurmountable before the advent of it.
Therefore we believe that it will reveal further truths concerning the
structure of space-time and the matter.

\section*{Acknowledgments}
We would like to thank A. Tsuchiya for the collaboration in the
early stage of this project. We also would like to 
thank D. Gross, Y. Makeenko, H. B. Nielsen and A. Polyakov for their 
valuable comments on our work.

\newpage
\appendix
\section{Fierz transformation} \label{sec:fierz}
\setcounter{equation}{0}

In this appendix, we investigate the totally symmetric tensors
which can be constructed out of a ten dimensional Majorana-Weyl
spinor $\xi$. 
The effective action
contains such a term as $trS^n$ where
$S^{\mu\nu}=\bar{\xi}\Gamma^{\mu\alpha\nu}\xi~{x_{\alpha}/ x^4}$
and $n$ is an even integer up to 8.
Therefore we are interested in totally symmetric tensors.

We first point out that
the only nonvanishing tensors which are quadratic in $\xi$ are
$\bar{\xi}\Gamma^{\mu\nu\lambda}\xi$.
Thus the Fierz transformation is performed quite easily, and we obtain
the following identities:
\beqa
(\bar{\xi}\Gamma^{\mu\alpha\beta}\xi)(\bar{\xi}\Gamma^{\nu}_{\;\;\alpha\beta
}\xi)
&=& 0,\label{eq:fierz1}\\
(\bar{\xi}\Gamma^{\mu\nu\alpha}\xi)
(\bar{\xi}\Gamma^{\lambda\rho}_{\;\;\;\;\alpha}\xi)
&=&
(\bar{\xi}\Gamma^{\mu\lambda\alpha}\xi)
(\bar{\xi}\Gamma^{\nu\rho}_{\;\;\;\;\alpha}\xi)
-(\bar{\xi}\Gamma^{\mu\rho\alpha}\xi)
(\bar{\xi}\Gamma^{\nu\lambda}_{\;\;\;\;\alpha}\xi)\label{eq:fierz2}.
\eeqa
From \eq{eq:fierz1} and \eq{eq:fierz2}, we can prove the following identity:
\beq
(\bar{\xi}\Gamma^{\mu\alpha}_{\;\;\;\;\beta}\xi)
(\bar{\xi}\Gamma^{\nu\beta}_{\;\;\;\;\gamma}\xi)
(\bar{\xi}\Gamma^{\lambda\gamma}_{\;\;\;\;\alpha}\xi)=0.\label{eq:fierz3}
\eeq
\underline{proof}:
Since the first factor of the left-hand side is antisymmetric in 
$\alpha$ and $\beta$,
the other factors can be antisymmetrized as 
the right-hand side of \eq{eq:fierz2}.
Therefore using \eq{eq:fierz2} we have
\beq
{\rm L.H.S.}
= -1/2 (\bar{\xi}\Gamma^{\mu\alpha}_{\;\;\;\;\beta}\xi)
(\bar{\xi}\Gamma^{\nu\lambda}_{\;\;\;\;\gamma}\xi)
(\bar{\xi}\Gamma^{\beta\;\;\gamma}_{\;\;\alpha}\xi),
\eeq
and it vanishes after the contractions of
$\alpha$ and $\beta$ due to \eq{eq:fierz1}. (q.e.d.)

Eq. (\ref{eq:fierz1}) shows that any totally symmetric tensor made of four
$\xi$ vanishes.
Since sixteen $\xi$ form Lorentz singlet,
we can see from the ``interchange of particles and holes''
that any totally symmetric tensor made of twelve $\xi$ must vanish.
From eq. (\ref{eq:fierz3}), we can also conclude the
absence of the totally symmetric tensors made of six and ten spinors in an
analogous way.

Thus we have shown 
that the totally symmetric tensors
made of $\xi$ are exhausted by the following list:
$\{1,\;
T^{\mu\nu\lambda\rho}=
(\bar{\xi}\Gamma^{\mu\alpha}_{\;\;\;\;\beta}\xi)
(\bar{\xi}\Gamma^{\nu\beta}_{\;\;\;\;\gamma}\xi)
(\bar{\xi}\Gamma^{\lambda\gamma}_{\;\;\;\;\delta}\xi)
(\bar{\xi}\Gamma^{\rho\delta}_{\;\;\;\;\alpha}\xi),\;
\xi^{16}\}$.
It is easily checked that $T^{\mu\nu\lambda\rho}$ is a totally symmetric 
traceless tensor in the following way.
Similar use of eqs. (\ref{eq:fierz2}) and (\ref{eq:fierz3})
to the above gives the antisymmetric part of $T^{\mu\nu\lambda\rho}$ as
\beqa
T^{\mu\nu[\lambda\rho]}
&=& -1/2
(\bar{\xi}\Gamma^{\mu\alpha}_{\;\;\;\;\beta}\xi)
(\bar{\xi}\Gamma^{\nu\beta}_{\;\;\;\;\gamma}\xi)
(\bar{\xi}\Gamma^{\lambda\rho}_{\;\;\;\;\delta}\xi)
(\bar{\xi}\Gamma^{\gamma\;\;\delta}_{\;\;\alpha}\xi),\n
&=& 0,
\eeqa
and from \eq{eq:fierz1} we immediately see that
$T^{\mu\nu\lambda}_{\;\;\;\;\;\;\lambda}=0$.
Therefore $T^{\mu\nu\lambda\rho}$ is totally symmetric and traceless.

\section{SU(2) Matrix Model} \label{sec:su2}
\setcounter{equation}{0}
In this appendix, we solve supersymmetric $SU(2)$  matrix models
in various dimensions, $D=3,4,6,10$:
\footnote{Some results presented
here
has been already reported in Ref. \cite{su2}. Similar calculations have also
been done in a different context in Ref. \cite{dbound}.  }
\begin{equation}
S  =  -{1\over g^2}Tr({1\over 4}[A_{\mu},A_{\nu}][A^{\mu},A^{\nu}]
+{1\over 2}{}^t{\psi}\alpha ^{\mu}[A_{\mu},\psi ]) ,
\end{equation}
where $ A_\mu (\mu=0,1,\cdots, D-1)$ and $ \psi$ are $ 2\times 2$
Hermite matrices, and  $\psi$ is a spinor in $D$ dimensional 
super Yang Mills theory.
Each spinor consists of $
p_D$ real components, where  $ p_D$ is 2,4,8, 16 for
$D$=3,4,6,10, respectively.
We denote $D$-dimensional alpha matrices by $ \alpha^\mu$'s
($ \alpha^\mu=\Gamma^0 \Gamma^\mu)$ and the first three of them can be
represented as follows:
\begin{eqnarray}
&&\alpha^i={\rho^i \otimes\mib{1}_{p_D/2}} ,\nonumber \\
&&\rho^i=
\cases{
\mib{1_2} & for $i$=0 \cr
\sigma^1 & for $i$=1\cr
\sigma^3& for $i$=2
}.
\end{eqnarray}

Performing the integral over fermionic variables gives the pfaffian
\footnote{Although the integration does not lead to a pfaffian in $D=6$,
the result (\ref{eq:eeaaa}) holds in this case as well.}
\begin{equation}
\hbox{Pf}( \alpha^\mu A_\mu)=\hbox{Pf}(\alpha^\mu T^a A^a_\mu),
\label{eq:pfaffian}
\end{equation}
where $ T^a$ is the generator of $SU(2)$ in the adjoint representation, namely
$ (T^a)_{bc}= \epsilon_{bac}$. Now we``rotate" $ A_\mu$ by Lorentz
transformation so that only $ A_0, A_1 $and $A_2$ are nonvanishing.
Eq. (\ref{eq:pfaffian}) then reduces to the three-dimensional calculation,
\begin{equation}
\bigl\{ \hbox{Pf} (\rho^i T^a A^a_i ) \bigr\}^{p_D/2} \sim
\bigl\{\epsilon_{ijk}\epsilon_{abc} A^a_i A^b_j A^c_k
\bigr\}^{p_D/2},
\label{eq:eeaaa}
\end{equation}
which corresponds to $(\det A )^{p_D/2}$ when we regard $ A^a_i$ as a 3$
\times$3 matrix.

Next step is the integration over three ten-dimensional vectors
$ A^a_\mu$, which
we reduce to the integration over three-dimensional vectors.
The Jacobian for this reduction is the volume of the parallelepiped spanned by
the three vectors, $ A^1, A^2$ and $ A^3$, to the $ (D-3)$-th,
which is nothing but
$ |\det A|^{D-3}$.

To estimate the behavior of the integral, we take the following parametrization
for three vectors $ A^a_i$,
\begin{equation}
\left (\matrix{
r \cr
\mib{0}\cr
}\right ),
 \left (\matrix{
a\cr
\mib{y}\cr
}\right ),
 \left (\matrix{
b\cr
\mib{z}\cr
}\right ),
\end{equation}
where $ \mib{y}$ and $ \mib{z}$ are two dimensional vectors while $ \mib{0}$ is
two dimensional zero vector.
After these considerations, we obtain
\begin{eqnarray}
&&\int dA d\psi e^{-S}\n
&=&\int dr r^2 da db d^2\mib{y}d^2\mib{z}
\bigl\{r(\mib{y}\times\mib{z)}\bigr\}^{p_D/2}|r(\mib{y}\times\mib{z)}|^{D-3}\n
&&\times \exp \bigl( -\frac{1}{g^2} \bigl[ r^2(\mib{y}^2+\mib{z^2})+
(a\mib{z}-b\mib{y})^2+(\mib{y}\times\mib{z})^2 \bigr] \bigr).
\label{eq:total}
\end{eqnarray}
Integrating over $ a$ and $ b$ yields
\begin{equation}
\int da db e^{-(a\mib{z}-b\mib{y})^2}
=\bigl(
\mib{z}^2\mib{y}^2-(\mib{z}\cdot\mib{y})^2 \bigr)^{-1/2}
=|\mib{y}\times\mib{z}|^{-1}.
\end{equation}

Next, let us integrate over $ \mib{y}$ and $ \mib{z}$.
For $D$=3 case the integral becomes identically zero,
since $\bigl\{r(\mib{y}\times\mib{z)}\bigr\}$ takes the both signs.
For the remaining dimensions,
we can estimate the integration
when $ r$ is large and small respectively,
which suffice the present purpose.
When $r$ is large, the first term of the argument of the exponential
in \eq{eq:total} becomes dominant.
We can perform the integrations over $ \mib{y}$ and $ \mib{z}$
by rescaling them as $ \mib{y}/r$ and $ \mib{z}/r$.
In this way eq. (\ref{eq:total}) is estimated to be
\beqa
\int dr r^{-(p_D/2+D-3)}
&=&\int r^{D-1} dr r^{-(p_D/2+2D-4)}\n
&=&
\cases{
 \int r^3dr\; r^{-6} & for $D$=4 \cr
\int r^5dr\; r^{-12} & for $D$=6\cr
\int r^9dr\; r^{-24} & for $D$=10}
\qquad (r^2 \gg g).
\label{eq:larger}
\eeqa
When $r$ is small, the first term of the argument of the exponential
becomes negligible.
We can perform the integrations over $ \mib{y}$ and $ \mib{z}$
independently from $r$. Eq. (\ref{eq:total}) is estimated
in this case to be
\begin{eqnarray}
\int dr r^{2+(p_D/2+D-3)}
&=&\int r^{D-1} dr r^{p_D/2}\n
&=& \cases{
\int r^3dr \; r^2 & for $D$=4 \cr
\int r^5dr \; r^{4} & for $D$=6\cr
\int r^9dr \; r^{8} & for $D$=10}
\qquad (r^2 \ll g).
\label{eq:smallr}
\end{eqnarray}

Finally we confirm that
the long distance behavior of $ r$ integral (\ref{eq:larger})
agrees with the result of the
perturbative calculation in the text.
Recall that
the one-loop effective action for $r$ and $\xi$ is written as
\beq
S_{\rm eff}^{1-{\rm loop}}[r,\xi] = tr \ln (1+S^{\mu\nu}),
\eeq
where
\beq
S^{\mu\nu} = \bar{\xi}\Gamma^{\mu\alpha\nu}\xi\frac{r_{\alpha}}{r^{4}}
\sim \xi^2/r^3.
\eeq
Thus the integration over $\xi$
\beq
\int d^Dr d^{p_D}\xi e^{-S_{\rm eff}[r,\xi]} 
\eeq
leads to \eq{eq:larger}.

\section{Branched polymer \label{app:BP}}
\setcounter{equation}{0}
In this appendix we give a brief introduction to thermodynamics of
branched polymers (BP) without self-avoiding effect in $D$ dimensions.
(see for example \cite{ambj}).
The partition function of BP with $N$ points and $(N-1)$ bonds
is given by
\begin{equation}
Z_N =  \sum_{\rm BP} \int \prod_{i=1}^{N-1} d^Dx^i
\prod_{(ij):{\rm  bond}}
f(x^i - x^j) \prod_{n=1}^{\infty} t_n^{\#n},
\label{BP10}
\end{equation}
where the summation is taken over all possible topologies of branched polymers.
$t_n$ denotes a positive weight assigned for the points
to which $n$ bonds are connected
and $ \#n$ denotes the number of such points in a given configuration.
The weight function for bonds $f(x)$ can be Fourier transformed as
\begin{equation}
\hat{f}(p) = \int d^Dx f(x) e^{-ipx} =  \hat{f}(0) (1-c_2 p^2 + c_4 p^4
+ \cdot   \cdot \cdot) ,
\label{BP20}
\end{equation}
with a positive coefficient $c_2$.
Since we are interested in the thermodynamic limit where $N$ goes to
infinity, we consider a grand canonical partition function:
\begin{equation}
Z = \sum_{N=1}^{\infty} N Z_N k_0^{(N-1)}.
\label{BP30}
\end{equation}
\par
\begin{figure}
\epsfxsize = 10 cm
\epsfysize =  3 cm
\centerline{\epsfbox{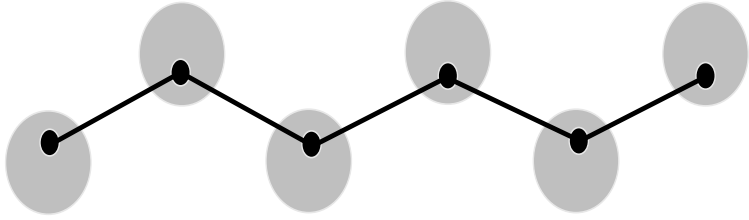}}
\caption{Graphical representation of a two point function.
A pair of points in a given configuration
can be connected by bonds in a unique way.
In this figure they are connected by five bonds.
Such bonds form a random walk type object.
The remaining points in a branched polymer
are connected to the points in this object.
They are represented by the blobs in this figure.}
\label{figA1}
\end{figure}

In order to demonstrate the scaling behavior,
we consider a two point correlation function.
If we pick up a pair of points in a particular configuration,
they are uniquely connected by $n$ bonds.
If we assume that they are separated by a fixed distance,
$n$ varies from a configuration to another.
Therefore
we can express a two point function in the momentum space
as follows:
\begin{eqnarray}
\hat{G}(p, k_0) &= &
\int d^D(x-y) G(x-y, k_0) e^{ip(x-y)} 
\nonumber \\
&\propto &
\sum_{n=1}^{\infty}k_0^{n+1} (t_1 +  t_2 b +  {t_3 \over 2!} b^2 + \cdot \cdot \cdot )^2
(\hat{f}(p))^n B^{n-1}.
\label{BP40}
\end{eqnarray}
A graphical illustration of this equation is given in fig.\ref{figA1}.
The factor $(t_1 +  t_2 b +   \cdot \cdot \cdot )$
is the contribution from each end  point in fig. \ref{figA1}.
$B$ represents each blob in fig. \ref{figA1}
 except  those at the end points.
\begin{figure}
\epsfxsize = 10 cm
\epsfysize =  3 cm
\centerline{\epsfbox{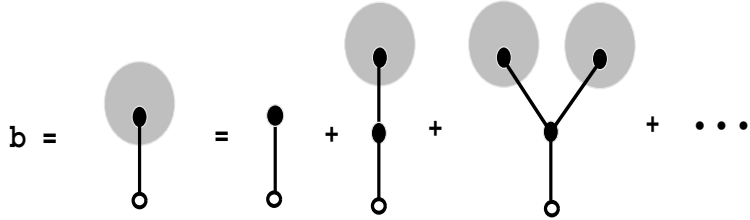}}
\caption{Graphical representation of the Schwinger Dyson equation for $b$. 
White circles is not associated with any wight.}
\label{figA2}
\end{figure}
$b$ is graphically represented in fig.\ref{figA2}.
It satisfies the following Schwinger Dyson equation
\begin{equation}
b= k_0 \hat{f}(0)( t_1 + t_2 b + {t_3 \over 2!} b^2+ \cdot \cdot \cdot)
= k h(b) ,
\label{BP50}
\end{equation}
where $k=k_0 \hat{f}(0)$.
The blob $B$ can be related to $h(b)$ as
\begin{equation}
B = t_2 + t_3 b + {t_4 \over 2!} b^2 + \cdot \cdot \cdot
= h'(b).
\label{BP60}
\end{equation}
From these considerations, we can see that $\hat{G}(p, k_0)$ can be written as
\begin{equation}
\hat{G}(p, k_0) \propto
{ k_0^2 h(b)^2 \hat{f}(p) \over 1 - k_0 \hat{f}(p)h'(b)}.
\label{BP70}
\end{equation}
We also note that $Z'(b) = h(b)$,
since the partition function (\ref{BP30}) is given by
\beq
Z(b) = 1 + t_1 b + t_2 b^2/2! + t_3 b^2 /3! + \cdot \cdot \cdot .
\label{BP75}
\eeq

\par
\begin{figure}
\epsfxsize = 5 cm
\epsfysize =  3 cm
\centerline{\epsfbox{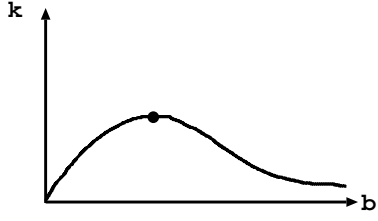}}
\caption{Typical relation between $k$ and $b$ is drawn.
A black dot in the figure indicates the critical point.
The large $N$ limit is
taken by approaching this point from the left.
}
\label{figA3}
\end{figure}

Since the averaged number of the points in the branched polymer
is given by
\begin{equation}
\bar{N} \sim {k \over Z} {\partial Z \over \partial k}
= {k h(b) \over Z(b) \left({\partial k \over \partial b}\right)},
\label{BP80}
\end{equation}
we must tune the fugacity $k_0$ so that $\partial k / \partial b $
approaches zero from a positive value.
Eq. (\ref{BP50}) solves
$k$ as a function of $b$: $k=b / h(b)$.
A typical solution is illustrated in fig.\ref{figA3}.
Generally $t_1 \ne 0$ and some of $t_n$'s  with $n$ larger than 2 are
non-zero. We observe that $k(b)$ vanishes at $b=0$ and $b=\infty$.
Since  it is positive definite, there is a critical
fugacity  $k_c$ where an averaged number becomes infinite.
\par
Near the critical point, $k(b)$ can be approximated by
\begin{equation}
k \sim k_c -{c \over 2} (b-b_c)^2 ,
\label{BP84}
\end{equation}
where $c$ is a positive constant determined by the set of $t_n$'s.
The partition function is given by the integral of $h(b)$ over $b$
and
it behaves near the critical point as
\begin{equation}
Z = {\rm const.} - {b_c \over k_c}\sqrt{{2(k_c -k) \over c}}.
\label{BP85}
\end{equation}
The universal part (the second term) determines the power
$-3/2$ in the following expansion:
\begin{equation}
Z = \sum_{N=1}^{\infty} N^{-3/2}  \left( {k \over k_c} \right)^N
\sim \sum_{N=1}^{\infty} N^{-3/2}  e^{ -{k_c-k \over k_c} N }.
\label{BP87}
\end{equation}
This is because the universal part of the infinite sum of the r.h.s.
in eq. (\ref{BP87}) can be estimated by an integral
$\int dN N^{-3/2} exp( - (k_c-k)N / k_c) $.
Therefore partition function for a fixed $N$ is
$Z_N= N^{-3/2} (\hat{f(0)} / k_c)^N $. The  leading
non-universal behavior is given by
\beq
Z_N \sim (\hat{f(0)} \alpha_c)^N ,
\eeq
where $\alpha_c =1/k_c$ takes a value of order one.
In a special case where $t_n=1$  for all $n$, we find that
$ k=b e^{-b}$ and
the critical values are $k_{c} = 1/e$ and $b_{c}=1$.
This is the case relevant to our analysis.
\par
Inserting $k h'(b)= 1- h(b) k'(b)$ (which is derived from eq.
(\ref{BP50})) into eq. (\ref{BP70}), we obtain a scaling
behavior of a two point function near the critical
point.
For small $p$, it behaves as
\begin{eqnarray}
\hat{G}(p, k_0) &\propto&
{ k_{0,c}^2 h^2(b_c)
\hat{f}(p) \over 1-  (\hat{f}(p) / \hat{f}(0))(1-\sqrt{2c(k_c-k)} h(b_c))    }
\nonumber \\
&\propto &
 {1  \over p^2 + M^2 \sqrt{(k_c-k)} }
\propto {1  \over  (p_{phy}^2 + M^2 )\sqrt{(k_c-k)} }
\label{BP100}
\end{eqnarray}
where $M^2 = \sqrt{2c}h(b_c) /c_2$.
In order to take a scaling limit, we have introduced the physical momentum
$p_{phy}= p (k_c-k)^{-1/4}$.
In this way, we have obtained the scaling relation:
\begin{equation}
(k_c-k)^{-1/2} G(x (k_c-k)^{-1/4}, k_0)
\stackrel{k \rightarrow k_c}{\longrightarrow} g(x)
\label{BP110}
\end{equation}
where $g(x)$ is a certain function.
Since the dominant contribution
comes from  $N \sim k_c/(k_c-k) $,
we can observe from eq. (\ref{BP110})  that the
Hausdorff dimension of a branched polymer is four.

\end{document}